\documentclass[sigconf]{acmart}

\usepackage{multirow}
\usepackage{colortbl}
\usepackage{subcaption}
\usepackage{rotating}

\definecolor{customblue}{RGB}{41,45,185}   
\definecolor{custompurple}{RGB}{249,95,0} 

\AtBeginDocument{%
  }

\setcopyright{cc}
\setcctype{by-nc-nd}
\copyrightyear{2026}
\acmYear{2026}
\acmDOI{10.1145/3772318.3790441}
\acmConference[CHI '26]{Proceedings of the 2026 CHI Conference on Human Factors in Computing Systems}{April 13--17, 2026}{Barcelona, Spain}
\acmBooktitle{Proceedings of the 2026 CHI Conference on Human Factors in Computing Systems (CHI '26), April 13--17, 2026, Barcelona, Spain}
\acmISBN{979-8-4007-2278-3/2026/04}

\begin{document}

\title{DuetUI: A Bidirectional Context Loop for Human-Agent Co-Generation of Task-Oriented Interfaces}

\author{Yuan Xu}
\orcid{0009-0004-0811-9505}
\affiliation{%
  \institution{The Hong Kong University of Science and Technology (Guangzhou)}
  \city{Guangzhou}
  \country{China}
}
\email{yxu712@connect.hkust-gz.edu.cn}

\author{Shaowen Xiang}
\orcid{0009-0000-0148-8253}
\affiliation{%
  \institution{The Hong Kong University of Science and Technology (Guangzhou)}
  \city{Guangzhou}
  \country{China}
}
\affiliation{%
  \institution{Shanghai Jiao Tong University}
  \city{Shanghai}
  \country{China}
}
\email{starryspace@sjtu.edu.cn}

\author{Yizhi Song}
\orcid{0009-0007-7648-8865}
\affiliation{%
  \institution{The Hong Kong University of Science and Technology (Guangzhou)}
  \city{Guangzhou}
  \country{China}
}
\email{ysong531@connect.hkust-gz.edu.cn}

\author{Ruoting Sun}
\orcid{0009-0007-7933-5338}
\affiliation{%
  \institution{The Hong Kong University of Science and Technology (Guangzhou)}
  \city{Guangzhou}
  \country{China}
}
\affiliation{%
  \institution{University of California, Berkeley}
  \city{Berkeley}
  \country{United States}
}
\email{linda123@berkeley.edu}

\author{Xin Tong}
\authornote{Corresponding author}
\orcid{0000-0002-8037-6301} 
\affiliation{%
  \institution{The Hong Kong University of Science and Technology (Guangzhou)}
  \city{Guangzhou}
  \country{China}
}
\affiliation{%
  \institution{The Hong Kong University of Science and Technology}
  \city{Hong Kong}
  \country{China}
}
\email{xint@hkust-gz.edu.cn}
\renewcommand{\shortauthors}{Xu et al.}

\begin{abstract}
    Large Language Models are reshaping task automation, yet remain limited in complex, multi-step real-world tasks that require aligning with vague user intent and enabling dynamic user override. From a \textbf{formative study} with 12 participants, we found that end-users actively seek to shape task-oriented interfaces rather than relying on one-shot outputs. To address this, we introduce the \textbf{human-agent co-generation paradigm}, materialized in \textbf{\textit{DuetUI}}. This LLM-empowered system unfolds alongside task progress through a \textbf{bidirectional context loop}---the agent scaffolds the interface by decomposing the task, while the user's direct manipulations implicitly steer the agent's next generation step. In a \textbf{technical ablation study} and a \textbf{user study} with 24 participants, \textbf{\textit{DuetUI}} improved task efficiency and interface usability, supporting more seamless human-agent collaboration. Our contributions include the proposal of this novel paradigm, the design of a proof-of-concept \textbf{\textit{DuetUI}} prototype embodying it, and empirical and technical insights from an initial evaluation of how this bidirectional loop may help align agents with human intent and inform future development.
\end{abstract}


\begin{teaserfigure}
  \includegraphics[width=\textwidth]{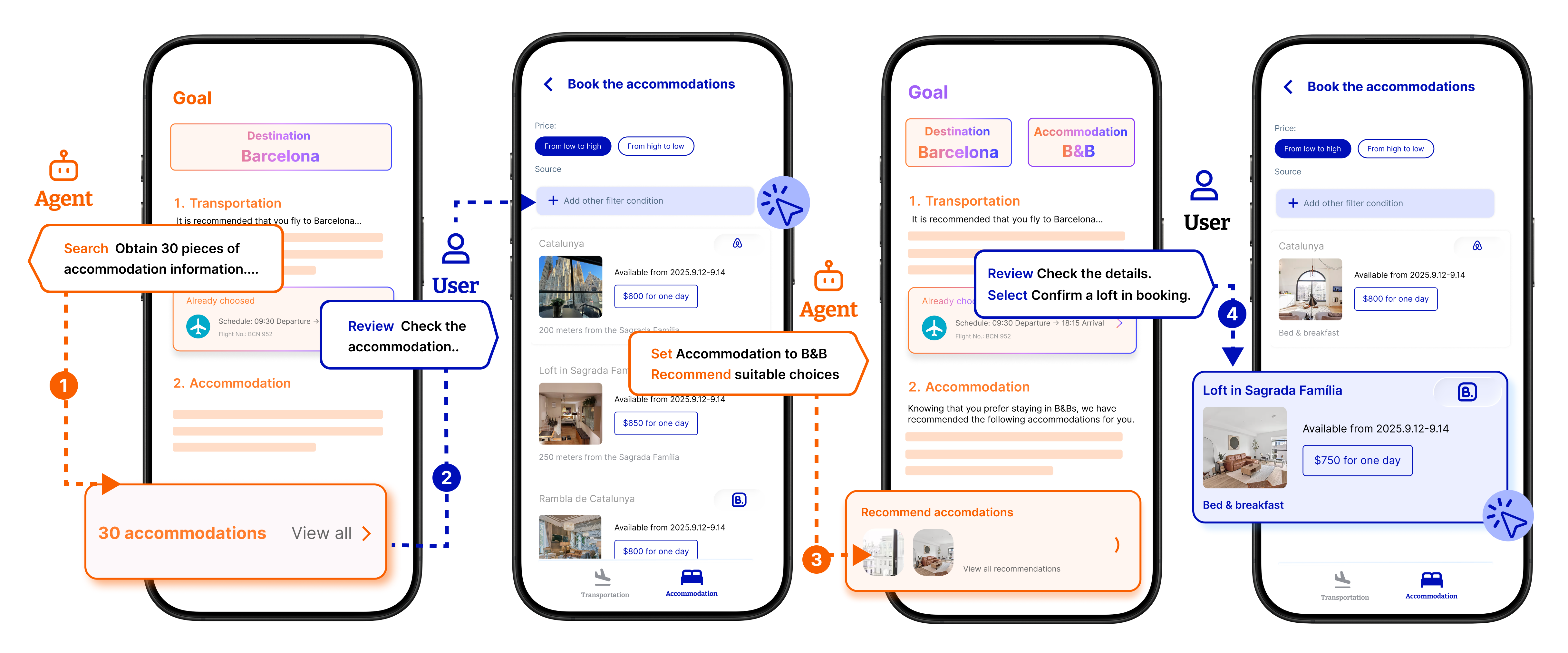}
  \caption{An illustration of our \textbf{human-agent co-generation paradigm}. The \textbf{bidirectional context loop} shows how agents and humans mutually steer the task. The agent decomposes a high-level goal into a concrete interface scaffold, while the user's direct manipulations on that interface provide implicit context that guides the agent's next step, enabling a fluid collaboration.} 
  \Description{An illustration of a human-agent co-generation paradigm presented across four sequential smartphone interface screens, demonstrating a bidirectional context loop between an agent and a user collaborating to plan accommodation for a trip to Barcelona. Starting with the leftmost screen (labeled Step 1): The screen displays the "Goal" (destination: Barcelona) with two subtasks (Transportation, marked as "Already Chosen," and Accommodation, in progress). The orange "Agent" module initiates action 1: "Search Obtain 30 pieces of accommodation information...", and the interface shows a prompt to "View all" 30 accommodations; the blue "User" label notes action 2: "Review Check the accommodation...". Moving to the second screen (Step 2): Titled "Book the accommodations," it lists three initial Barcelona accommodations (e.g., "Catalunya" for \$600/day, 200 meters from Sagrada Família) with price sorting and filter options; the agent’s orange label indicates action 3: "Set Accommodation to B\&B Recommend suitable choices," responding to the user’s review. The third screen (Step 3) updates the "Goal" section to show "Accommodation" tagged as "B\&B," with the agent’s module noting: "Knowing that you prefer staying in B\&B, we have recommended the following accommodations for you," and the user’s blue label outlines action 4: "Review Check the details. Select Confirm a loft in booking." The rightmost screen (Step 4) finalizes the loop: It displays the user-selected "Loft in Sagrada Família" (a B\&B option priced at \$750/day, available 2025.9.12-14), highlighting the agent’s adaptation to the user’s implicit context (from initial 30 options to a targeted B\&B recommendation) and the user’s direct manipulation to confirm a specific choice—embodying the fluid collaboration described in the figure caption, where the agent decomposes high-level goals into interface scaffolds, and the user’s interface interactions guide the agent’s subsequent steps.}
  \label{fig:teaser}
\end{teaserfigure}


\maketitle

\section{Introduction}
Large Language Models (LLMs) are shaping recent developments in task automation, leveraging their powerful multimodal understanding and planning capabilities to transition from predefined, script-based methods to intelligent GUI (Graphical User Interface) agents~\cite{zhangLargeLanguageModelBrained2025,qinUITARSPioneeringAutomated2025, songVisionTaskerMobileTask2024, jiangAppAgentXEvolvingGUI2025}. These agents can interpret natural language commands and interact with GUI to perform tasks~\cite{songVisionTaskerMobileTask2024, qinUITARSPioneeringAutomated2025}. However, their efficacy often diminishes when confronted with complex, multi-step daily workflows that span multiple services~\cite{liuMobileStewardIntegratingMultiple2025, zhangLargeLanguageModelBrained2025, songVisionTaskerMobileTask2024}. The low success rates of fully autonomous agents in such scenarios stem from an important gap between the end-user's often ambiguous or evolving intent~\cite{zamfirescu-pereiraWhyJohnnyCant2023} and the agent's actions, suggesting the value of human-in-the-loop approaches. These approaches allow end-users to manually configure workflows~\cite{yinOperationCognitionAutomatic2025} or intervene at critical steps~\cite{huangPrompt2TaskAutomatingUI2025,yeInteractionIntelligenceDeep2025}. Despite improving task outcomes, these methods are still largely agent-centric and often do not fully support a fluid partnership, often presenting interfaces that are difficult to learn and demanding to operate for end-users~\cite{horvitz1999principles,hoqueVisualizationHumanCenteredAI2024}.

An emerging paradigm seeks to generate task-oriented~\cite{yehHowGuideTaskoriented2022, lewis1993task} user interface, shifting the focus from full automation to human-agent collaboration. This task-oriented UI generation paradigm produces a unified, coherent interface that aggregates the necessary functionalities for the end-user, rather than having an agent opaquely execute a task from start to finish~\cite{zhaoSurveyLargeLanguage2025}.
Unlike prior work on UI generation aimed at designers~\cite{swearnginScoutRapidExploration2020, chenICardGenerativeAISupported2025} and engineers~\cite{luMistyUIPrototyping2025, maDynExDynamicCode2025},
this paradigm targets end-users who lack technical expertise~\cite{geGenComUIExploringGenerative2025, caoGenerativeMalleableUser2025}. However, this shift introduces its own set of unanswered questions rooted in long-standing HCI (Human-Computer Interaction) challenges. How can a generated interface be aligned with an end-user's mental model to bridge the gulf of execution and evaluation~\cite{normanDesignEverydayThings2013,morrisHCIAGI2025}? How can an end-user's vague, evolving intent be reconciled with an agent's need for structured instructions~\cite{zamfirescu-pereiraWhyJohnnyCant2023,morrisHCIAGI2025,vaithilingamDynaVisDynamicallySynthesized2024}? Critically, how can we design for a seamless balance between automation and human control~\cite{horvitz1999principles,cuiNoRightOnline2023,endsleyHereAutonomyLessons2017,beaudouin-lafonGenerativeTheoriesInteraction2021}, allowing users to shift between delegation and direct manipulation~\cite{yeInteractionIntelligenceDeep2025}?

To address these challenges, we explore a new paradigm: \textbf{human-agent co-generation}, materialized in a system \textbf{\textit{DuetUI}}. The system consists of a \textbf{bidirectional context loop}, where the agent and the end-user collaboratively shape the task and the interface in a process of man-computer symbiosis~\cite{lickliderManComputerSymbiosis1960, kayEarlyHistorySmalltalk1996}. The agent scaffolds the process by decomposing the task into a manageable interface, while the end-user's direct manipulations on that interface implicitly guide the agent's subsequent actions. We investigate the following research questions to explore this approach:
\begin{itemize}
    \item \textbf{RQ1}: How can LLMs be leveraged to support a collaborative process where agents and end-users co-generate a malleable and task-oriented interface for complex daily tasks?
    \item \textbf{RQ2}: How does such a co-generated interface, mediated by a bidirectional context loop, affect system performance, objective interaction behavior, and the overall human-agent collaboration experience?
\end{itemize}

To answer these questions, we first conducted a formative study with 12 participants. Through a think-aloud protocol, we observed how end-users interact with and attempt to drive generative, task-oriented interfaces. Our study uncovered four recurring challenges: end-users reveal their intent gradually, struggle with opaque AI behaviors, entangle task goals with interface needs, and desire fluid collaboration where control shifts between them and the system. These findings highlight two central tensions: (1) the intertwining of task and interface and (2) the balance between end-user control and AI automation, which directly informed the bidirectional context loop design in \textit{DuetUI} and echoed established guidelines for human-AI interaction~\cite{amershiGuidelinesHumanAIInteraction2019, horvitz1999principles}. We then conducted a technical evaluation to assess the system’s generative quality and functional alignment. This evaluation used a hybrid approach: automated LLM-as-judge assessment for functional equivalence matching, combined with expert manual scoring across efficiency, layout, and completeness. Finally, we ran a user study with 24 participants comparing \textit{DuetUI} to a baseline, analyzing objective interaction logs and subjective user experience. Results show \textit{DuetUI} significantly improved task efficiency and interface usability, enabling more seamless human-agent collaboration. Our contributions are threefold:
\begin{itemize}
    \item Design insights from a \textbf{formative study} that reveals how non-expert end-users approach, drive, and interact with generative task-oriented interfaces in everyday workflows.
    \item \textbf{\textit{DuetUI}}, a novel system that embodies bidirectional \textbf{human-agent co-generation paradigm}. This is achieved through a \textbf{bidirectional context loop} that enables a fluid workflow where agent-led task decomposition and user-driven manipulations mutually shape the collaborative process.
    \item Technical and empirical findings from a \textbf{multi-dimensional evaluation}, demonstrating how the bidirectional context loop improves objective task performance and interaction efficiency while fostering superior user satisfaction compared to existing paradigms.
\end{itemize}

\section{Related Work}
\subsection{Paradigms in Task Automation}
The methodology for accomplishing complex tasks has evolved through several distinct paradigms, each showing a different relationship between the user and the computer. The primitive paradigm is direct manipulation, where the user mentally plans and sequentially executes each step on a GUI (Figure~\ref{fig: Related Work}a). While intuitive, this approach becomes a bottleneck for large-scale, repetitive, or multi-service workflows~\cite{zhangLargeLanguageModelBrained2025, yinOperationCognitionAutomatic2025,songVisionTaskerMobileTask2024,caoGenerativeMalleableUser2025}. To address these limitations, the field has shifted towards full automation (Figure~\ref{fig: Related Work}b). While early approaches relied on rigid, scripted actions~\cite{ivancicRoboticProcessAutomation2019}, recent LLM-driven GUI agents like UI-Tars~\cite{qinUITARSPioneeringAutomated2025} and CogAgent~\cite{hongCogAgentVisualLanguage2024} introduce significantly greater dynamism by translating natural language directly into operations. However, this agent-centric paradigm remains insufficient for real-world scenarios, where user goals are not only often ambiguous~\cite{zamfirescu-pereiraWhyJohnnyCant2023,normanDesignEverydayThings2013} but also evolve continuously during the process~\cite{stanfordonlineStanfordSeminarCreating2023,wuAIChainsTransparent2022}, leaving a persistent gap between user intent and agent actions.

Recognizing this rigidity, a third paradigm of human-in-the-loop supervision has been explored (Figure~\ref{fig: Related Work}c-d). This brings the user back into the process, but typically in a supervisory role. The interaction is often turn-based, with users and agents taking sequential control~\cite{huangPrompt2TaskAutomatingUI2025, yeInteractionIntelligenceDeep2025}, or it requires users to configure a structured plan like a visual workflow upfront~\cite{yinOperationCognitionAutomatic2025, zhongHelpVizAutomaticGeneration2021}. While an improvement, this paradigm maintains a rigid separation that creates significant challenges for effective collaboration. From the perspective of Norman's theory, the turn-based model widens the ``gulf of execution''~\cite{normanDesignEverydayThings2013}, forcing users to translate their fluid goals into discrete instructions. This approach conflicts with HCI guidelines that emphasize continuous alignment~\cite{morrisHCIAGI2025} and support for ``small tweaks''~\cite{vaithilingamDynaVisDynamicallySynthesized2024}, causing the system to miss granular needs—the ``last mile'' of personalization that remains inaccessible without active human engagement.

\subsection{User Interface Generation}
Parallel to task automation, user interface generation seeks to enhance human-computer interaction by computationally creating and adapting interfaces~\cite{jiangComputationalApproachesUnderstanding2022,gajosSUPPLEAutomaticallyGenerating,wobbrockAbilityBasedDesignConcept2011}. Pioneering work in this area used optimization-based approaches, with systems like SUPPLE~\cite{gajosSUPPLEAutomaticallyGenerating} and UNIFORM~\cite{nicholsUNIFORMAutomaticallyGenerating2006} formalizing layout design as combinatorial optimization tasks to improve user performance and experience~\cite{oulasvirtaCombinatorialOptimizationGraphical2020}. The advent of large-scale GUI datasets such as RICO~\cite{dekaRicoMobileApp2017} and MUD~\cite{fengMUDLargeScaleNoiseFiltered2024} enabled the subsequent shift to data-driven methods, which employed deep learning models like GNNs~\cite{jiangGraph4GUIGraphNeural2024} and Transformers~\cite{jingLayoutGenerationVarious2023} to learn from existing designs.

More recently, LLM-driven UI generation has emerged, demonstrating strong zero-shot capabilities~\cite{geGenComUIExploringGenerative2025, wuUICoderFinetuningLarge2024}. However, due to inherent stability and consistency issues~\cite{caoGenerativeMalleableUser2025}, its primary application remains confined to serving as an auxiliary tool under the supervision of professionals, such as developers and designers. In this role, LLMs aid in asset search~\cite{parkLeveragingMultimodalLLM2025}, concept validation~\cite{swearnginScoutRapidExploration2020}, and prototyping~\cite{luMistyUIPrototyping2025}, thereby improving professional workflow efficiency. While these technologies are becoming a viable medium for human-AI collaboration~\cite{yenMemoletReifyingReuse2024, caoGenerativeMalleableUser2025}, a critical limitation persists: no matter how efficient professionals become, they cannot scale to tailor designs for every unique user need, yet users typically lack the expertise to design for themselves. Furthermore, even professionally produced designs impose a significant learning curve on end-users. Drawing inspiration from pattern languages in UI design~\cite{alexanderPatternLanguageTowns1977, tidwellDesigningInterfacesPatterns2010}, we propose moving beyond single-task generation to create a coherent and persistent task-oriented workspace. This approach aims to enforce structural consistency across a workflow while still allowing for fluid, malleable interaction at each step.

\subsection{Toward Human-Agent Co-Generation}
\begin{figure}[h!]
    \centering
    \includegraphics[width=\linewidth]{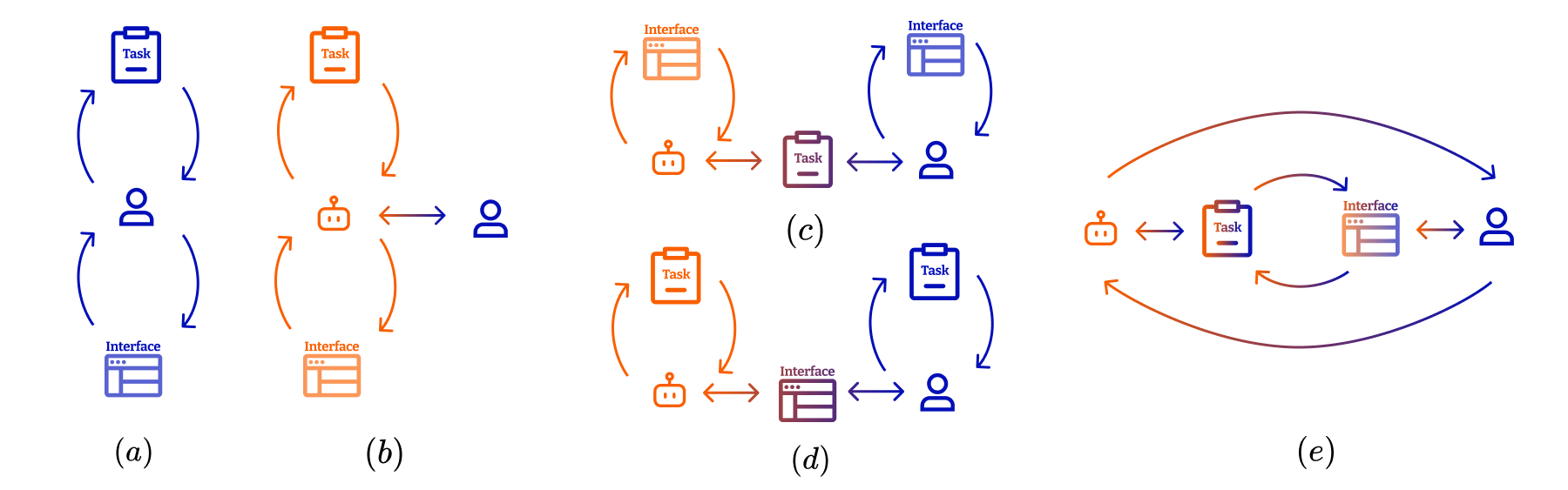}
    \caption{The Evolution of Task Automation Paradigms. (a) Traditional direct manipulation by the user. (b) Agent-centric full automation, which excludes the user. (c-d) Recent human-in-the-loop approaches that treat the user as a supervisor. (e) Our proposed bidirectional co-generation paradigm, enabling a seamless human-agent partnership.}
    \Description{This figure illustrates the evolution of task automation paradigms, shown as a sequence from left to right: (a) Direct manipulation: The user performs tasks manually through traditional interfaces. (b) Agent-centric automation: The agent operates fully automatically, with the user excluded from the process. (c–d) Human-in-the-loop approaches: The agent performs most actions, while the user supervises, corrects, or approves decisions. (e) Bidirectional co-generation (DuetUI’s paradigm): The user and agent work as equal partners in a continuous loop, jointly shaping both tasks and interfaces.The figure contrasts earlier paradigms that either burden the user or exclude them, with the proposed co-generation model that emphasizes seamless human–agent collaboration.}
    \label{fig: Related Work}
\end{figure}

The limitations in current task automation and UI generation converge on a central disconnect between human intent and AI execution. To address this, we propose a paradigm of \textit{human-agent co-generation} (Figure~\ref{fig: Related Work}e). This concept draws heavily from the vision of End-User Development (EUD), which distinguishes itself from \textit{Participatory Design} by empowering users to adapt software artifacts during actual use rather than solely at design time~\cite{barricelliEnduserDevelopmentEnduser2019a}. EUD has evolved significantly from early demonstration systems like Vegemite~\cite{lin2009end} to recent approaches like ``Vibe Coding''~\cite{sarkar2025vibecodingprogrammingconversation} and LLMs-driven assistance~\cite{gaoEasyAskInAppContextual2024}. These advances consistently make users active participants in software creation rather than passive consumers. However, substantial barriers remain in effectively realizing this partnership. Traditional EUD approaches often introduce a steep learning curve or operational burden~\cite{caoGenerativeMalleableUser2025} while modern LLM-based tools frequently struggle because users provide incomplete raw requirements due to a lack of domain knowledge~\cite{zhangEmpoweringAgileBasedGenerative2024}. Furthermore, while recent algorithmic research introduces bidirectional deliberation mechanisms~\cite{zhang2024enhancinglanguagemodelrationality, tran2023exploring}, these typically rely on the agent's internal forward and backward reasoning loops. Such models often relegate the human to the passive role of an explicit evaluator or prompter and neglect the potential for the user's natural operational behavior to serve as a continuous guidance signal.

We address these disconnects through a \textit{bidirectional context loop} where the human and agent co-construct both the interface and task. Unlike previous methods that only built on explicit programming, our approach interprets the user's direct manipulation of the co-generated interface as \textit{implicit guidance} for the agent's subsequent actions. This mechanism creates a tighter feedback loop rooted in the vision of man-computer symbiosis~\cite{lickliderManComputerSymbiosis1960, kayEarlyHistorySmalltalk1996}. By making the interface the shared medium of communication, we aim to bridge the gulfs of execution and evaluation~\cite{normanDesignEverydayThings2013}, allowing users to seamlessly alternate between delegation and control.

\section{Formative Study}
\subsection{Method}
To explore a human-agent co-generation paradigm, we conducted a formative study to understand how end-users currently engage with agent and interface to address their daily task. We recruited 12 participants (8 female, aged 20-27) via social media to identify tensions and challenges in their collaborative processes.

\subsubsection*{Design Probe System}
We developed a web-based probe system (detailed in Appendix~\ref {section: probe}) to facilitate the design sessions, serving both as a guide and a creative toolkit that allowed non-expert participants to articulate ideas in multiple ways. The system integrated four key components: an \textbf{chatbot} (\textit{Gemini}) that answered questions and performed web searches to support information gathering, an \textbf{UI prototyper} (\textit{dev0}\footnote{\url{https://v0.app/docs/api/platform}}) that translated natural language into interactive mockups for iterative refinement, a \textbf{notepad} for documenting thoughts and detailing desired features or interaction flows, and a \textbf{drawing canvas} for sketching layouts and other visual concepts.

\subsubsection*{Tasks}
Participants were asked to design an interface for two complex, real-world tasks. Drawing from established HCI research~\cite{caoGenerativeMalleableUser2025, yenMemoletReifyingReuse2024, yinOperationCognitionAutomatic2025} and mobile app datasets~\cite{dekaRicoMobileApp2017, fengMUDLargeScaleNoiseFiltered2024}, we selected \textbf{Task 1: Travel Planning} and \textbf{Task 2: Relocating to a New City}. These tasks were selected as they reflect daily, multi-step goals that are not solved by any single application and could be performed individually, allowing us to focus purely on the design activity.

\subsubsection*{Procedure}
Each study session was conducted remotely via video conference and lasted approximately 60 minutes. All sessions were screen- and audio-recorded with prior informed consent from the participants. This study and the subsequent research were approved by the Institutional Review Board under protocol number HKUST(GZ)-HSP-2025-0181. The procedure for each session was as follows, as shown in Figure~\ref {fig:Procedure}:
\begin{enumerate}
    \item \textbf{Introduction And Consent:} We began by introducing the study's objectives, walking the participant through the functionalities of the design probe system, and signing the consent form.
    \item \textbf{Task 1 (Unprompted Design):} For the first task (Travel Planning), participants recalled and described past experiences, focusing on process and pain points. They then used the probe system to design their ideal interface following a \textbf{think-aloud protocol}, creating and refining designs while verbalizing rationales for UI choices, features, and interaction flows.
    \item \textbf{Task 2 (Stimulated Design):} The second task (Relocating) followed a similar procedure, but with one key addition. Before beginning their design, participants were shown several short demo videos of existing task automation interfaces~\cite{qinUITARSPioneeringAutomated2025,caoGenerativeMalleableUser2025}. They were asked to provide feedback, which then served as a stimulus for their own design process.
    \item \textbf{Semi-Structured Interview:} We concluded with a semi-structured interview to reflect on the session. We asked participants about the design strategies they employed, the difficulties they encountered, and their overall perception of designing solutions for complex cross-application tasks.
\end{enumerate}

\subsubsection*{Data Collection and Analysis}
We collected several forms of qualitative data for each participant: screen and audio recordings, all digital artifacts created (UIs, notes, sketches), and audio recordings from the concluding interview. In total, 10.41 hours of recordings were gathered (not including the introduction section). Our analysis involved two researchers independently coding this data. They subsequently met to consolidate their codes and synthesize the results, forming the basis for the findings discussed below.

\subsection{Formative Study Findings and Design Goals}
\label{sec:formative_study}
Our analysis of the formative study revealed four key findings that illuminate the central tensions within the current paradigm of human-agent collaboration. From these insights, we derived a set of design goals aimed at fostering a more fluid and co-generative system.

\subsubsection{[F1] Intent is Emergent and Incrementally Refined Through Interaction}
Our formative study revealed that user intent is not a static, upfront specification but rather an emergent property of the interaction process. All participants (12/12) externalized their goals in a phased manner, beginning with what they described as a ``\textit{basic motivation}'' (P4-6) to generate an initial result. This baseline served as a crucial artifact for evaluation, allowing them to ``\textit{see how far it is from the final requirement, and then make modifications on this basis}'' (P9). This interaction, in turn, reshaped their intent. As P11 noted, ``\textit{after I see something, I might have different ideas.}'' Participants conceptualized this workflow as a series of ``\textit{interconnected}'' steps (P2, P5, P7, P9), which P9 aptly termed a ``\textit{superposition}'' that makes the result \textit{increasingly precise.}''

\begin{table}[h!]
\centering
\caption{A Six-Stage Model of Human-Agent Co-Generation Derived from Formative Study Findings.}
\label{tab:staged_co_generation}
\renewcommand{\arraystretch}{1.5} 
\begin{tabular}{c p{0.15\linewidth} p{0.70\linewidth}} 
\hline
\textbf{No.} & \textbf{Stage} & \textbf{Goal} \\ \hline
1 & \textbf{Define} & Capture the user's high-level goal to establish the core task. \\
2 & \textbf{Empathize} & Elicit further details and constraints to deepen the understanding of user needs. \\
3 & \textbf{Plan} & Structure the user's requirements into a coherent plan composed of multiple subtasks. \\
4 & \textbf{Explore} & For a given subtask, gather external data and explore potential solutions. \\
5 & \textbf{Refine} & Adjust the details and parameters of a proposed solution for a subtask. \\
6 & \textbf{Duet} & Finalize and execute the task through iterative refinement cycles. \\ \hline
\end{tabular}
\Description{The table presents a conceptual model for human-agent co-generation, outlining six distinct stages and their corresponding goals. The stages are: Define, Empathize, Plan, Explore, Refine, and Duet. This model illustrates a structured progression from capturing high-level user intent to achieving final task execution through iterative collaboration.}
\end{table}

This finding directly challenges the paradigm of one-shot generation and led us to our first design goal: \textbf{[DG1] Support a continuous and collaborative co-creation process}. Drawing on theories of situated action~\cite{bardram1997plans}, which frame plans as adaptable resources rather than rigid scripts, our goal was to design an interface that functions not as a final output, but as a dynamic scaffold. This scaffold should facilitate an ongoing dialogue where the user and agent collaboratively navigate the task from initial ideation to final execution. To structure this dialogue, we synthesized our observations with established frameworks in design thinking~\cite{brown2009change,deMul1996exploration} to propose a conceptual model for this process. This model, detailed in Table~\ref{tab:staged_co_generation}, breaks down the co-generative workflow into six distinct stages.

\subsubsection{[F2] The Opaque AI Created a Gulf of Execution for Users.}
A majority of participants (11/12) faced a significant gulf of execution~\cite{normanDesignEverydayThings2013}\footnote{The ``gulf of execution'' is a concept in human-computer interaction that describes the gap between a user's intention (what they want to achieve) and the actions required by the system to execute that intention. A wide gulf means the user struggles to figure out how to operate the system to accomplish their goals.}, stemming from a poor mental model of the AI's capabilities. Users perceived the AI as a ``\textit{black box}'' (P10), expressing doubts about whether it could comprehend their instructions (P1-3, P5-6, P8). This uncertainty led them to avoid certain modalities entirely, such as the drawing canvas, because they felt they were ``\textit{not good at sketching}'' (P2) and lacked confidence in the AI's interpretive power. To bridge this gap, users proposed solutions like learning from examples (P2-5) and direct manipulation of the output (P2-3, P6).

To address this, our second design goal was: \textbf{[DG2] To achieve effortless instrumentality and bridge interaction gulfs}. Drawing from Shneiderman's principles of direct manipulation~\cite{shneiderman1983direct}, we aimed to empower users with a sense of direct control and minimal cognitive load. This meant providing a continuous representation of the task objects and enabling rapid, reversible actions. The system must make its capabilities and intentions transparent through clear signifiers and feedforward~\cite{normanDesignEverydayThings2013}, ensuring that users can readily understand what is possible and how to execute their goals without needing specialized knowledge in UI design or prompt engineering.

\subsubsection{[F3] User Expressions Entangled Tasks and Interfaces.}
While users could describe a desired outcome, they struggled to decompose high-level goals into structured, machine-executable components. Their expressions often intertwined task needs with interface expectations, as seen with P2: ``\textit{I want to first search for travel guides on social media [task], browse and bookmark some posts [interface], then summarize a rough plan [task]...}'' However, these expressions were frequently underspecified. For example, the same user assumed a``bookmark'' function would implicitly include a``delete'' option, stating ``\textit{if you can add, you must also be able to remove}'' (P2), a detail the system needed to infer.

This challenge motivated our third design goal: \textbf{[DG3] To enable task-oriented interface unfolding}. This goal is grounded in the methodology of task-centered design~\cite{lewis1993task} and other prior work~\cite{yehHowGuideTaskoriented2022,caoGenerativeMalleableUser2025}, which advocates for making users' work and goals the primary driver of interface structure. Instead of generating generic UIs, the system should construct interfaces that are fundamentally organized around the user's task. By decomposing the task's semantic structure and mapping it directly to the UI hierarchy and layout, the system ensures the interface is always a relevant and direct medium for task accomplishment.

\subsubsection{[F4] Users Desire a Dynamic, Mixed-Initiative Collaboration.}
All participants (12/12) expressed a preference for workflows where control could shift fluidly between them and the AI. As P2 explained, ``\textit{At the beginning, I don't really know what I'm doing, so I need the system to recommend... But once I have a plan, I should take the lead, and the AI should become a helper.}'' This highlights a need for continuous, context-aware collaboration rather than rigid modes of either pure delegation or direct manipulation. Users wanted the AI to integrate information while leaving judgment to them (11/12) and valued its ability to store preferences, making the interaction feel personalized instead of with a system that is ``\textit{missing a `me'}'' (P1).

This desire for a fluid partnership shaped our final goal: \textbf{[DG4] To foster mutual awareness for shared autonomy}. This goal extends beyond simple turn-taking to embrace the principles of mixed-initiative interaction~\cite{horvitz1999principles}, where either the human or the agent can seize the initiative to advance the shared goal. To achieve this, the system aim to facilitate the establishment of common ground~\cite{clark1991grounding} by making the actions and inferences of both parties mutually transparent. Drawing on the concept of ``shared autonomy''~\cite{cuiNoRightOnline2023,endsleyLevelAutomationForms2018,endsleyHereAutonomyLessons2017}, control can be handed off and taken back fluidly, with any operation by one party instantly reflected in the other's context.

\section{DuetUI: A System for Human–Agent Co-Generation}

To materialize our proposed paradigm of human-agent co-generation, we drew upon insights from our formative study and theoretical foundations. We posit that this paradigm hinges on a dual-layered mechanism: at the surface, it enables fluid control switching; at a deeper level, it requires that the context generated in each turn be mutually perceived and synchronized to ensure progressive momentum. We developed \textbf{\textit{DuetUI}} to embody these principles through four synergistic features, collectively establishing what we term the \textbf{Bidirectional Context Loop}.

\subsection{System Features}
\label{sec:features}

We operationalize this loop through four features designed to function as a cohesive whole. To guarantee the progressive momentum we envision, \hyperref[feat:staged_generation]{\textbf{\textit{Staged Co-Generation}}} first imposes a structural path for the interaction. Within this structure, we design \hyperref[feat:tangible_agency]{\textbf{\textit{Tangible Agency}}} and the \hyperref[feat:action_history]{\textbf{\textit{Bidirectional Action History}}} as complementary mechanisms for intent analysis: the former provides an intermediate language to clarify \textit{explicit} intent, while the latter records behavioral data to capture the \textit{implicit} intent latent in user actions. Finally, to ensure these captured intents are correctly grounded, \hyperref[feat:duality]{\textbf{\textit{Task-Interface Duality}}} aligns human-agent understanding across corresponding levels of granularity.

\subsubsection{Staged Co-Generation}
\label{feat:staged_generation}
To fulfill our \textbf{[DG1]}, DuetUI eschews the ``one-shot'' generation paradigm. Instead, it operationalizes the six-stage conceptual framework (Table~\ref{tab:staged_co_generation}) derived from our research. This Staged Co-Generation mechanism structures human-agent interaction as a progressive, iterative dialogical process. As elaborated in the Usage Scenario (Section~\ref{section: usage example}), the system incrementally guides users through discrete phases—ranging from initial goal formulation to final execution refinement—with each stage featuring an interface specifically tailored to its unique collaborative objectives.

\begin{figure*}[!ht]
    \centering
    \includegraphics[width=\linewidth]{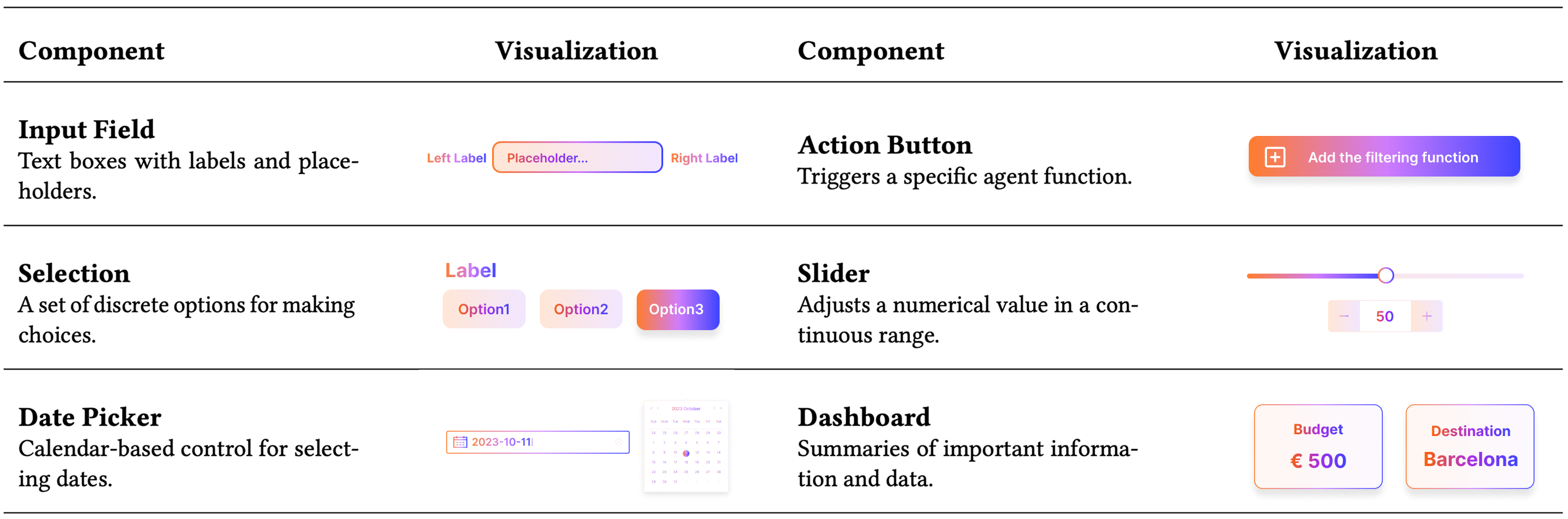}
    \caption{The suite of UI components that provide Tangible Agency, allowing users to directly manipulate the agent's capabilities.}
    \Description{This table lists the UI components that enable tangible agency in DuetUI. Each component is paired with a short description and a visualization example showing how it appears in the interface.Input fields – Allow users to type or provide free-form input. The visualization shows a text box where the user can enter information.Selections – Provide predefined options to choose from, such as dropdown menus or checkboxes. The visualization depicts a menu with multiple choices.Action buttons – Trigger specific operations, like “Search” or “Book.” The visualization shows a clickable rectangular button with a label.Sliders – Enable adjustment of values along a continuous range, such as budget or time. The visualization shows a horizontal bar with a movable handle.Date pickers – Support choosing specific dates or ranges. The visualization shows a calendar layout where the user selects a day.Dashboards – Summarize multiple pieces of information at once, such as prices, reviews, or recommendations. The visualization shows a panel combining charts or lists.}
    \label{tab:tangible_agency_components}
\end{figure*} 

\subsubsection{Tangible Agency}
\label{feat:tangible_agency}
To address \textbf{[DG2]}, we move beyond conventional conversational interfaces by introducing the concept of \textbf{Tangible Agency}. This design principle makes an agent's abstract capabilities concrete and directly manipulable, transforming the agent from a conversational partner into a set of tangible instruments the user can wield. By externalizing these capabilities, it empowers users to explicitly articulate their needs and exert precise control over the underlying intelligence. Instead of relying on ambiguous textual prompts, users interact with it through a dynamic set of UI controls embedded within the task document. This approach renders the agent's functions visible and actionable, bridging the gulf of execution. The core components that enable this tangible interaction are detailed in Figure~\ref{tab:tangible_agency_components}.

\begin{table}[ht!]
  \centering
  \caption{The Task--Interface Duality.}
  \label{tab:duality_map}

  \renewcommand{\arraystretch}{1.3} 
  \setlength{\tabcolsep}{6pt} 
  \begin{tabular}{ p{0.46\linewidth}  p{0.46\linewidth} }
    \toprule

    \multicolumn{1}{c}{\textbf{Task Decomposition}} & 
    \multicolumn{1}{c}{\textbf{Interface Description}} \\
    \midrule

    {\color{custompurple}\textbf{Task}} \newline
    \includegraphics[width=\linewidth]{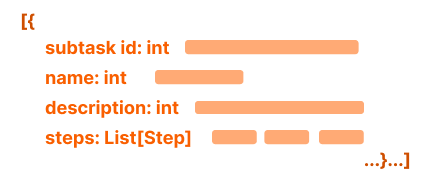} 
    & 
    {\color{customblue}\textbf{Navigation}} \newline
    \includegraphics[width=\linewidth]{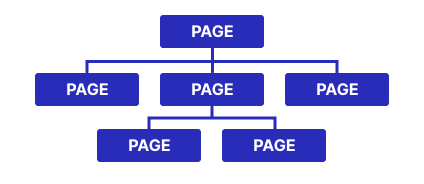} \\

    {\color{custompurple}\textbf{Subtask}} \newline
    \includegraphics[width=\linewidth]{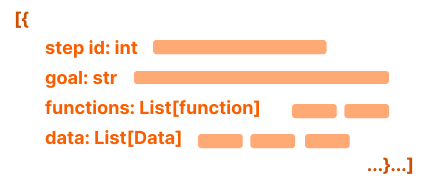}
    & 
    {\color{customblue}\textbf{Page}} \newline
    \includegraphics[width=\linewidth]{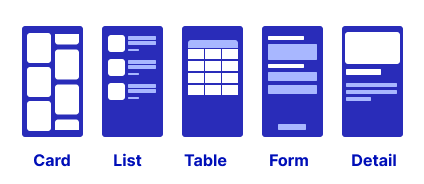} \\

    {\color{custompurple}\textbf{Data}} \newline
    \includegraphics[width=\linewidth]{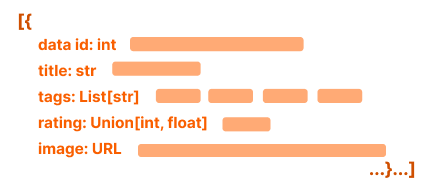}
    & 
    {\color{customblue}\textbf{Component}} \newline
    \includegraphics[width=\linewidth]{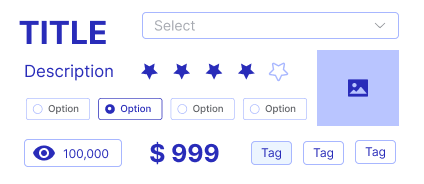} \\
    \bottomrule

  \end{tabular}

  \Description{A comparison table with two main columns. The left column represents Task Decomposition with purple labels (Task, Subtask, Data) positioned above their corresponding diagrams. The right column represents Interface Description with blue labels (Navigation, Page, Component) positioned above their corresponding diagrams.}
\end{table}

\subsubsection{Task-Interface Duality}
\label{feat:duality}
Embodying \textbf{[DG3]}, DuetUI establishes a rigorous semantic duality between the logical task structure and Interface representation. This principle ensures that every component of the task plan is instantiated as a directly manipulable interface element. By strictly coupling abstract logic with concrete UI controls, the system facilitates precise human-AI alignment across multiple levels of hierarchy, allowing users to synchronize with the agent's reasoning from high-level goals down to granular actions (Table~\ref{tab:duality_map}).

\subsubsection{Bidirectional Action History}
\label{feat:action_history}
\begin{figure}[h]
    \centering
    \includegraphics[width=0.95\linewidth]{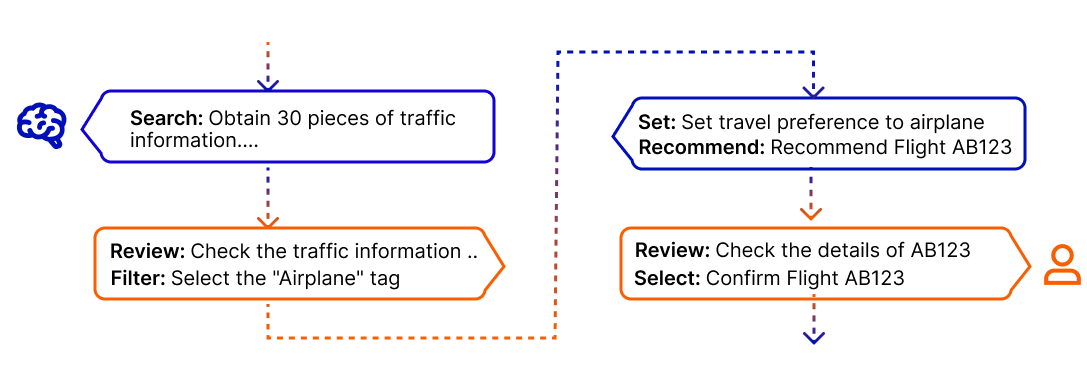} 
    \caption{An example of the Bidirectional Action History. The log shows a sequence of agent actions and user actions, enabling the agent to infer intent and collaborate effectively.}
    \Description{The figure presents an example of the Bidirectional Action History. The figure is shown as a vertically ordered log timeline, recording both agent and user actions in sequence. The log begins with the agent searching and filtering traffic information. Next, the agent sets the travel preference to “airplane.” The agent then recommends a specific flight. Finally, the user reviews and confirms Flight AB123. In the diagram, each step is listed line by line, alternating between agent actions and user actions, forming a combined history. The visual emphasizes that by referencing this shared log of actions, the agent can better infer user intent and collaborate more effectively.}
    \label{fig:action_history}
\end{figure}

To achieve the mutual awareness outlined in \textbf{[DG4]}, DuetUI relies on a \texttt{Bidirectional Action History}. This shared log captures every meaningful action from both the agent and the user. Agent actions are implicitly communicated through visible changes in the UI, while user actions (e.g., clicks, inputs) are explicitly recorded. This history becomes the living context for collaboration. For example, as shown in Figure~\ref{fig:action_history}, the agent's action of searching for transport options is displayed in the UI. The user's subsequent action of clicking a filter tag for ``airplane'' is logged. The agent immediately processes this implicit intent, refines its search to favor flights, and recommends \textit{Flight AB123}. The user's final selection and confirmation are also captured, completing a full cycle of transparent, history-aware interaction.

\subsection{Example Usage Scenario}
\label{section: usage example}
To better demonstrate the core features of DuetUI, we present a scenario where a user, Harry, plans a trip to Barcelona, from initial ideation to completing bookings.

\textbf{Define.} Harry wants to travel to Barcelona. Initially, his intention is vague, with only a destination in mind. He opens DuetUI and enters a simple prompt: ``I want to go to Barcelona for a trip''\includegraphics[height=2.5ex]{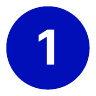}. DuetUI parses this broad request and generates an interface to clarify his intent, offering potential travel types to choose from, such as ``Family Vacation,'' ``Romantic Getaway,'' or ``Solo Exploration.'' \includegraphics[height=2.5ex]{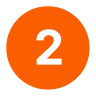}. Harry selects the option that best fits his plan: ``I want to take a solo trip to Barcelona.'' \includegraphics[height=2.5ex]{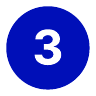}.

\begin{figure}[h!]
    \centering
    \includegraphics[width=0.42\linewidth]{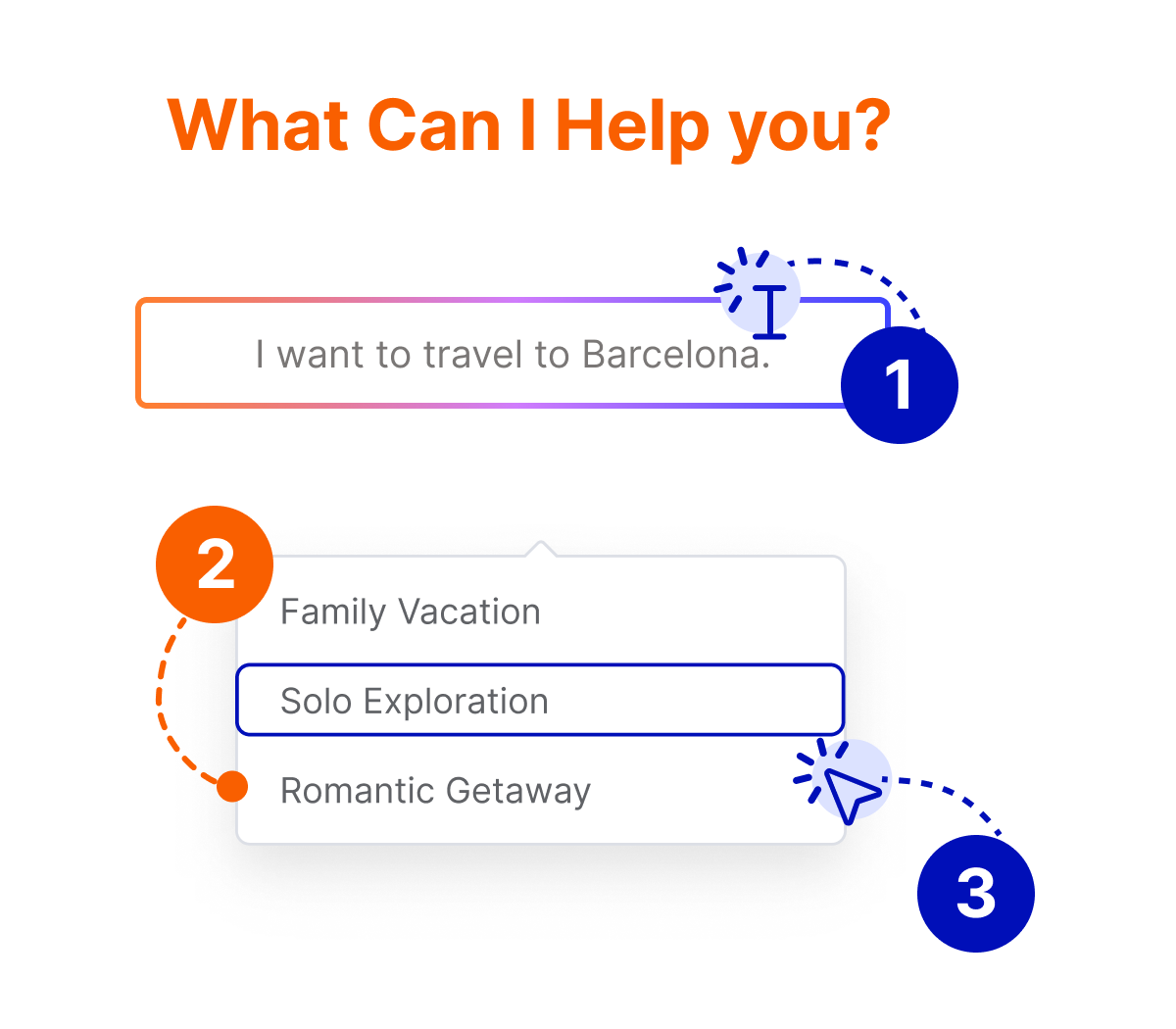}
    \caption{Define}
    \Description{The figure shows an illustration of the Define stage. The user types in the input field: “I want to travel to Barcelona.” Below, the system presents several trip type options, such as Family Vacation, Romantic Getaway, and Solo Exploration.}
    \label{fig:define}
\end{figure}

\textbf{Emphasize.} Once Harry clarifies his travel type, DuetUI begins to help him specify the details through a series of questions. The system asks, ``How would you like to travel to Barcelona?'' providing options for ``Flight,'' ``Train,'' or ``Car,'' while also inquiring about his approximate budget \includegraphics[height=2.5ex]{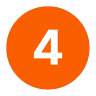}. Harry chooses ``Flight'' but is unsure about the budget yet, so he skips that question for now \includegraphics[height=2.5ex]{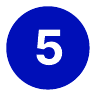}. Based on this initial dialogue, DuetUI consolidates the information into a structured user profile interface. This view includes the confirmed destination, transportation choices, and more granular preferences like flight class or accommodation requirements\includegraphics[height=2.5ex]{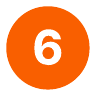}. Harry quickly edits his profile on this interface, for instance, selecting ``Economy Class'' for his flight\includegraphics[height=2.5ex]{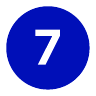}.

\begin{figure}[ht!]
    \centering
    \includegraphics[width=0.85\linewidth]{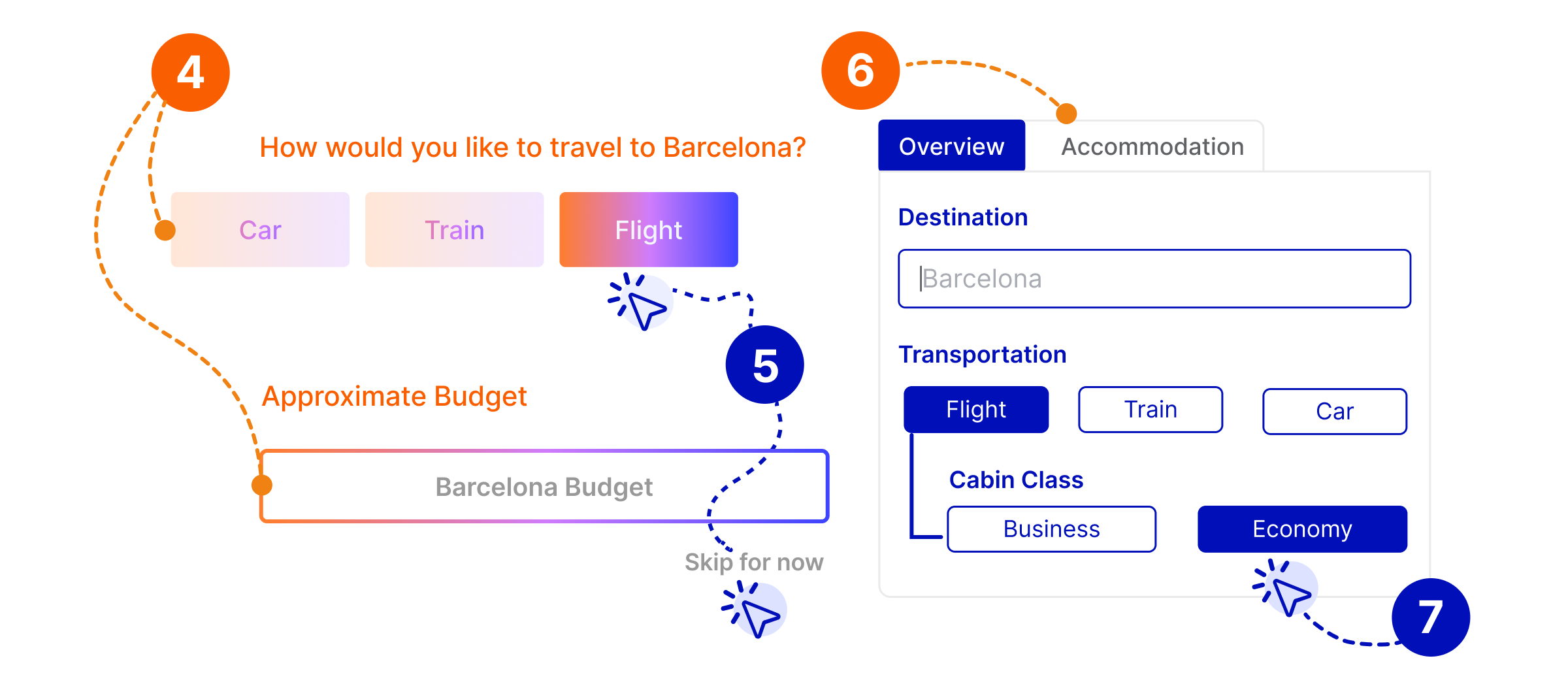}
    \caption{Emphasize}
    \Description{The figure illustrates the Empathize stage. The user selects Flight as the preferred transportation method. A budget for the Barcelona trip is entered. The system displays an overview form summarizing destination and transportation options. Within this form, the user chooses Economy cabin class.}
    \label{fig:Emphasize}
\end{figure}

\textbf{Plan.} After capturing Harry's basic requirements, DuetUI automatically generates a clear task plan for his trip. The plan is presented as a linear sequence of cards, breaking down the process into subtasks like ``Book round-trip flights,'' ``Plan daily itinerary,'' and ``Find and book accommodation.''\includegraphics[height=2.5ex]{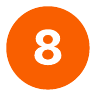} Harry decides it's better to secure his lodging before planning the daily activities, so he drags and drops the ``Find and book accommodation'' card to a position before ``Plan daily itinerary.''\includegraphics[height=2.5ex]{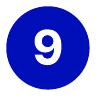}

\begin{figure}[ht!]
    \centering
    \includegraphics[width=0.42\linewidth]{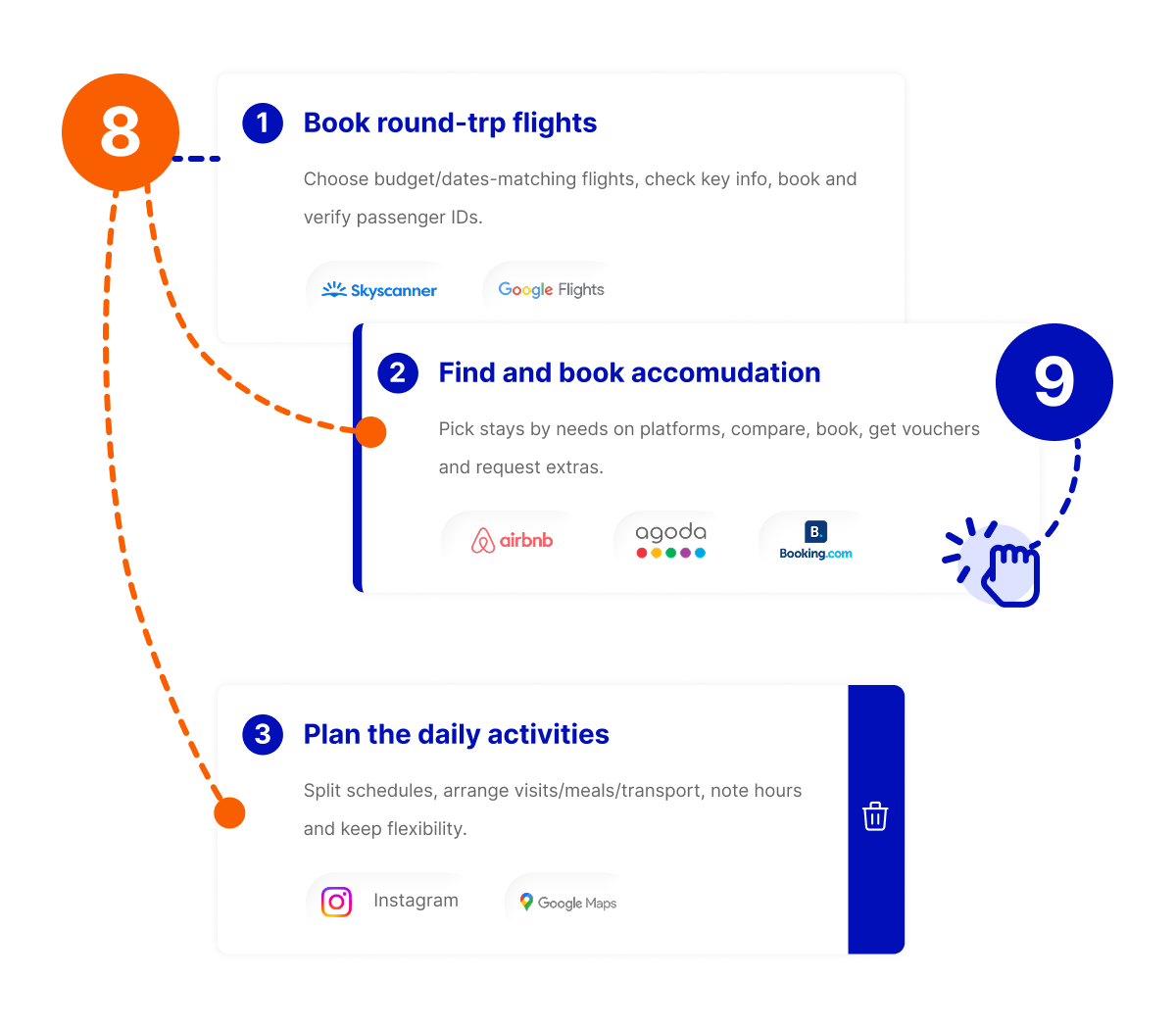}
    \caption{Plan}
    \Description{The figure illustrates the Plan stage. The interface shows a step-by-step checklist, each with numbers and icons. Steps include: Book round-trip flights (with a Skyscanner and Google Flights icon). Find and reserve accommodations (with an Airbnb, agoda and Booking icon). Plan daily activities (with a Google Maps and Instagram icon).}
    \label{fig:Plan}
\end{figure}

\textbf{Explore.} Upon receiving Harry's reordered task structure, DuetUI proceeds with the first task: exploring flight and accommodation options. The system proactively searches the web for data matching his initial preferences and presents a consolidated view of preliminary flight and hotel suggestions \includegraphics[height=2.5ex]{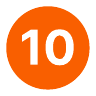}. Reviewing these initial options gives Harry a clearer idea of the potential costs. He decides to set his budget to around €1000 and specifies a preference for a local experience in an Airbnb. He adjusts the budget slider to €1000 and adds ``Airbnb'' as an accommodation preference \includegraphics[height=2.5ex]{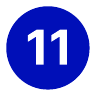}. 

\begin{figure}[ht!]
    \centering
    \includegraphics[width=0.85\linewidth]{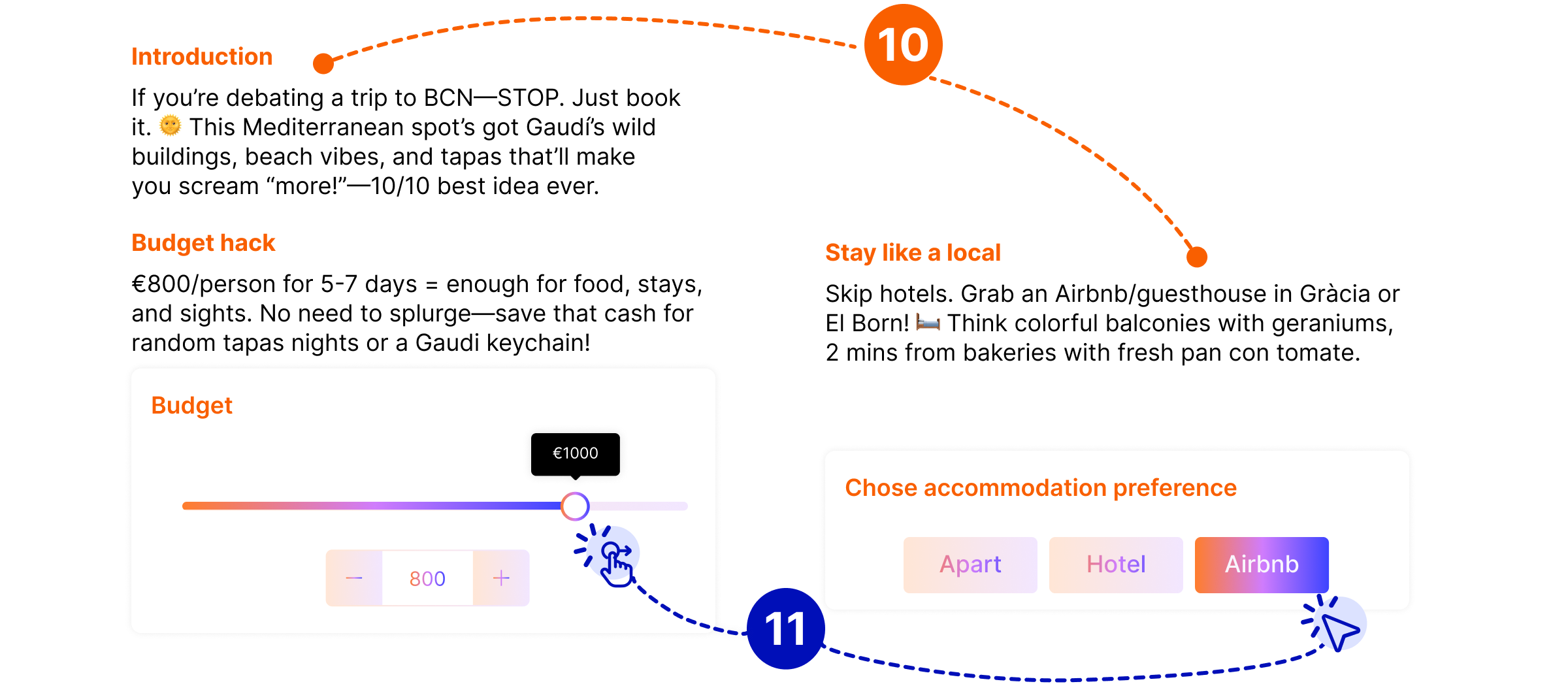}
    \caption{Explore}
    \Description{The figure illustrates the Explore stage. The system provides an introduction to Barcelona, including budget tips and local stay suggestions. A budget slider is shown, which the user sets to €1000. The user selects Airbnb as the preferred accommodation type.}
    \label{fig:Explore}
\end{figure}

\textbf{Refine.} In response to Harry's updated preferences, DuetUI immediately refreshes the search results and provides an interactive interface equipped with filtering and sorting capabilities \includegraphics[height=2.5ex]{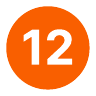}. Harry begins to browse the recommended Airbnbs. To make the best choice, he opens the detailed view, sorts the listings by ``Price: Low to High,'' and uses the map filter to view only properties near the ``Sagrada Família.'' After comparing a few options, he selects and books his preferred one \includegraphics[height=2.5ex]{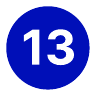}. 

\begin{figure}[ht!]
    \centering
    \includegraphics[width=0.42\linewidth]{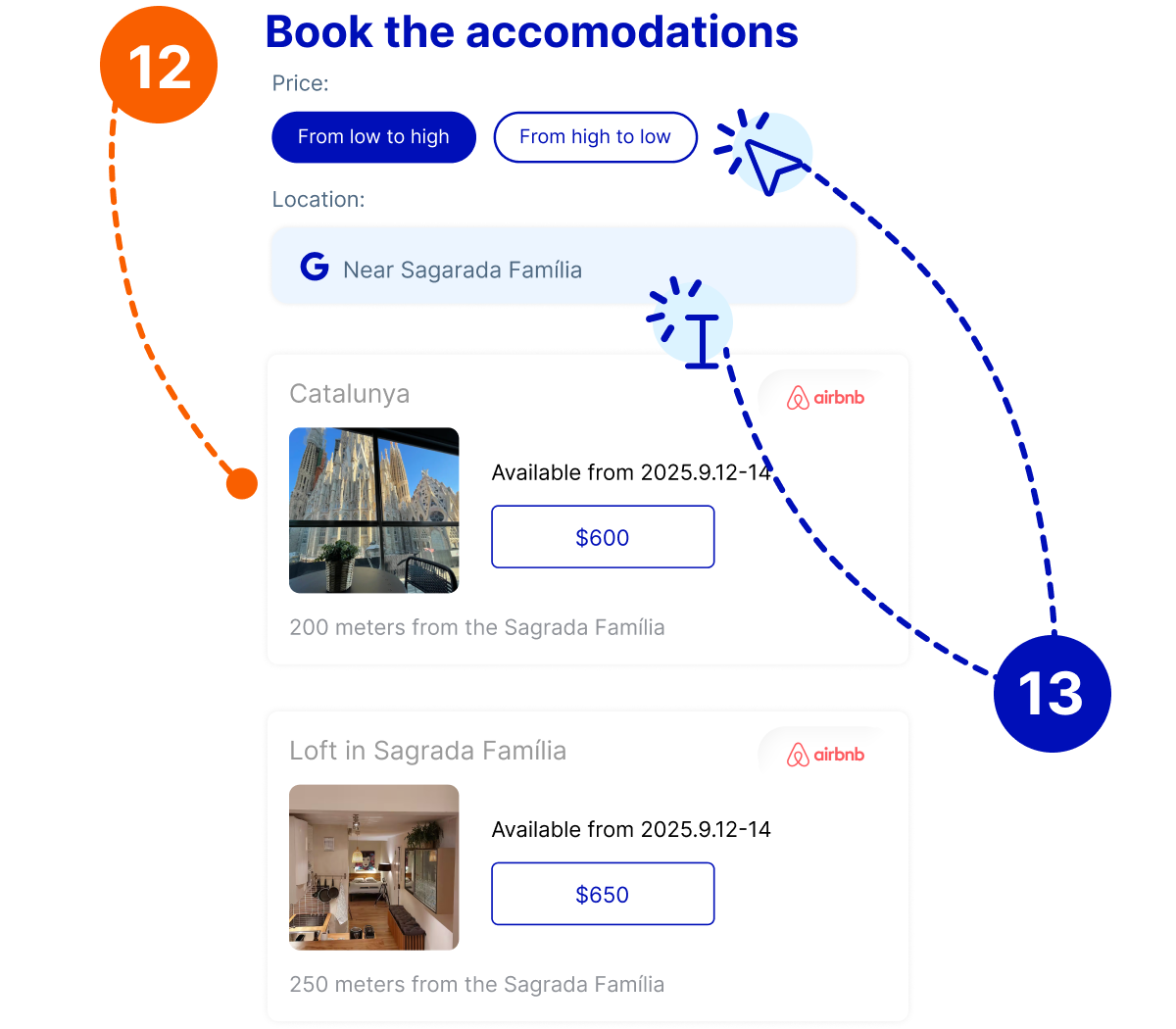}
    \caption{Refine}
    \Description{Illustration of the Refine stage: the user books accommodations by filtering by price and location near Sagrada Família, then compares two Airbnb listings, Catalunya at \$600 and Loft in Sagrada Família at \$650.}
    \label{fig:Refine}
\end{figure}

\textbf{Duet.} While Harry was booking his accommodation, DuetUI actively analyzed his behavior. It observed his focus on properties near the Sagrada Família and his use of price sorting. The system inferred that the Sagrada Família is a key point of interest for him and that he is a budget-conscious traveler. Based on this, after he completes the booking, DuetUI proactively pushes a detailed guide to the Sagrada Família. It also recommends other nearby, affordable cultural attractions, clearly indicating their distance from his Airbnb and their ticket prices \includegraphics[height=2.5ex]{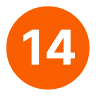}. Harry is pleased with these personalized recommendations, browses the guides, and favorites several points of interest \includegraphics[height=2.5ex]{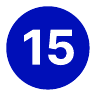}. Finally, DuetUI integrates Harry's booked flight, his Airbnb, and his favorite attractions to automatically generate a comprehensive and well-structured travel itinerary for his trip to Barcelona \includegraphics[height=2.5ex]{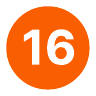}. 

\begin{figure}[ht!]
    \centering
    \includegraphics[width=0.85\linewidth]{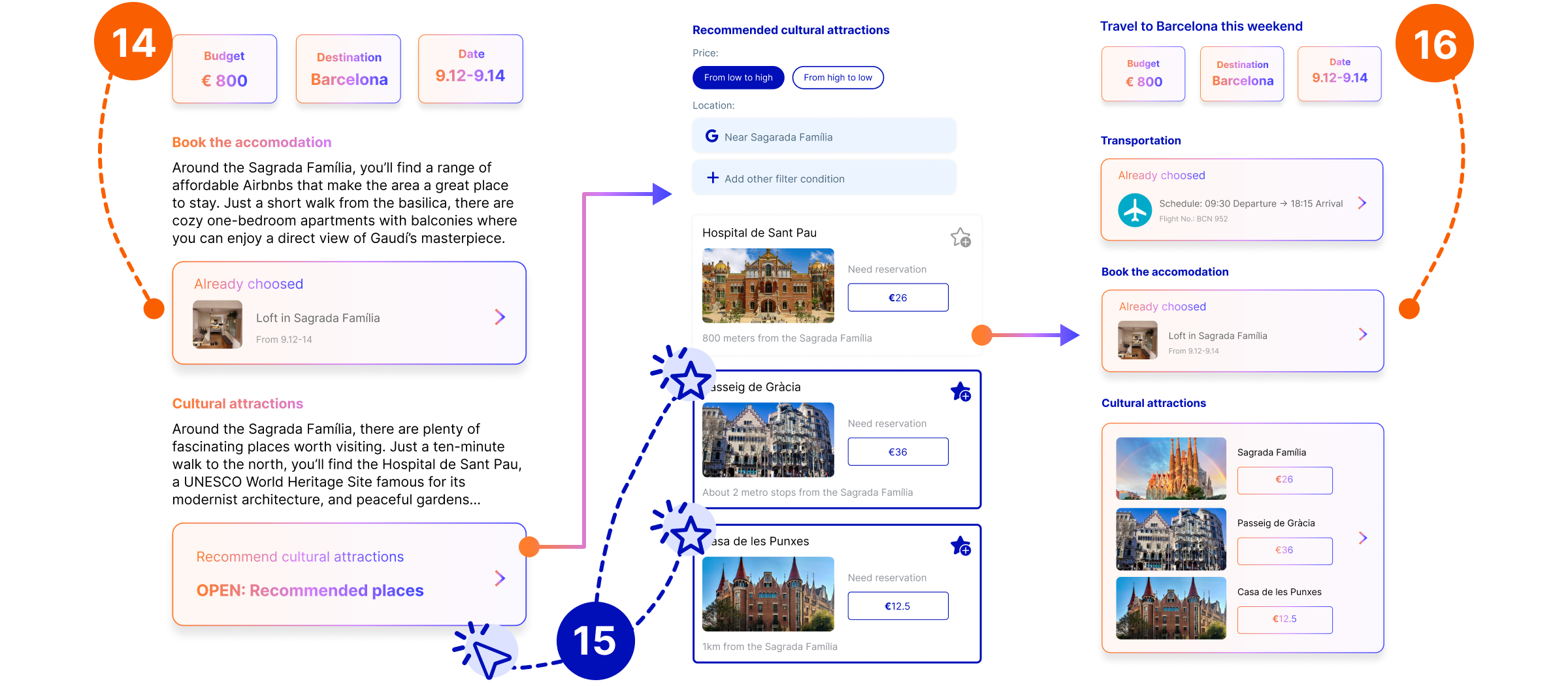}
    \caption{Duet}
    \Description{Illustration of the Duet stage: users confirm their budget, destination, and dates, select accommodation near Sagrada Família, and receive recommendations for cultural attractions such as Hospital de Sant Pau, Passeig de Gràcia, and Casa de les Punxes, which are integrated with transport and booking details.}
    \label{fig:Duet}
\end{figure}

\section{Implementation}
\begin{figure*}[ht]
    \centering
    \includegraphics[width=0.95\linewidth]{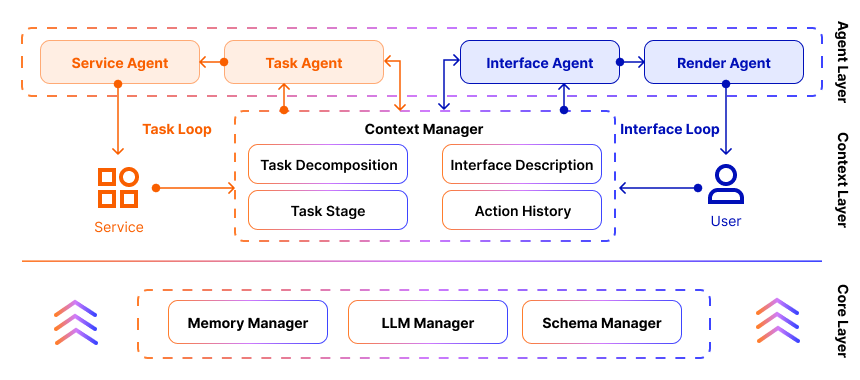}
    \caption{The DuetUI System Architecture. The diagram illustrates the three-tier structure (Core, Context, and Agent Layers) and the two operational loops (Task and Interface) that drive the interaction. The central Context Layer acts as the hub, mediating the state between the agent's internal logic and the user-facing interface to enable the bidirectional context loop.}
    \Description{The figure shows the DuetUI system architecture organized into three horizontal layers. At the top, the Agent Layer contains four linked modules: Service Agent, Task Agent, Interface Agent, and Render Agent, forming a pipeline from services to the user interface. In the middle, the Context Layer centers on a Context Manager with four blocks: Task Decomposition and Task Stage on the task side, and Interface Description and Action History on the interface side. Dashed arrows labeled Task Loop and Interface Loop connect the agents with the Context Manager and a user icon on the right. At the bottom, the Core Layer provides shared infrastructure via Memory Manager, LLM Manager, and Schema Manager. Together, the arrows indicate how agents, context, and core services interact continuously.}
    \label{fig:architecture}
\end{figure*}

To materialize our proposed human-agent co-generation paradigm, we designed and implemented \textbf{\textit{DuetUI}} as a system capable of sustaining a continuous, symbiotic dialogue between a user and an AI agent. The central challenge was engineering an architecture that could operationalize the core mechanism of this paradigm: the \textbf{bidirectional context loop}. Our solution, visualized in Figure~\ref{fig:architecture}, is a three-layer architecture built around a central \textbf{Context Layer}, which acts as the shared ``common ground'' for collaboration. This is supported by a \textbf{Core Layer} for foundational services and an \textbf{Agent Layer} for AI-driven logic. Together, these layers orchestrate two interconnected cycles---the \textbf{Task Loop} and the \textbf{Interface Loop}---which together realize the fluid, back-and-forth interaction central to our paradigm.

\subsection{System Architecture}

\subsubsection{Core Layer}
The Core Layer provides the fundamental, stateless services supporting the entire application. It includes three key components: the \texttt{MemoryManager} for information persistence, which assigns a session ID to each task and manages data in a key-value store; the \texttt{LLMManager}, which handles all interactions with LLM endpoints (including OpenAI GPT-4o \footnote{\url{https://openai.com/}} and Groq LLaMA3-70B \footnote{\url{https://groq.com/}} ), managing prompt assembly, network requests, retry logic, and JSON parsing; and the \texttt{SchemaManager}. As DuetUI is a multi-agent system, the \texttt{SchemaManager} is critical for enforcing data consistency, using Pydantic\footnote{\url{https://github.com/pydantic/pydantic}} schemas to define unified data structures across all agents.

\subsubsection{Context Layer}
The Context Layer is the system's stateful core, orchestrated by the central \texttt{ContextManager}. This manager maintains the complete shared state of the human-agent interaction, serving as the single source of truth that enables the operational loops. Its primary components directly reflect our core features: it tracks the current \texttt{TaskStage}, which dictates the phase of the \textit{Staged Co-Generation} process (see Section~\ref{feat:staged_generation}). It maintains the \texttt{Task-Interface Duality}, the crucial semantic link between the logical \texttt{TaskDecomposition} and the visual \texttt{InterfaceDescription} (see Section~\ref{feat:duality} and is detailed within the attached materials). Most importantly, it compiles the \texttt{Bidirectional Action History}, a comprehensive log of every tangible user manipulation and agent-initiated operation, which provides the rich, moment-to-moment context for collaboration (see Section~\ref{feat:duality} and Section~\ref{feat:action_history}).

\subsubsection{Agent Layer}
\label{subsubsection: Agent Layer}
The Agent Layer consists of four specialized agents designed to act upon the state maintained in the Context Layer. The \texttt{TaskAgent} is responsible for maintaining and modifying the \texttt{TaskDecomposition}, a three-level hierarchical structure (task, subtask, data). Concurrently, the \texttt{InterfaceAgent} manages the \texttt{InterfaceDescription}, a parallel structure representing the UI (navigation, page, components). The \texttt{ServiceAgent} executes calls to external services (simulated via LLMs, and are detailed within the attached materials) to fulfill data requirements defined in the task plan. Finally, the \texttt{RenderingAgent} translates the \texttt{InterfaceDescription} into a fully interactive web interface using the Element Plus \footnote{\url{https://github.com/element-plus/element-plus}} library and Vue.js framework \footnote{\url{https://github.com/vuejs/core}}.

\subsection{The Bidirectional User-Agent Collaboration Pipeline through Task Loops and Interface Loops} Building upon the static definitions within the Context Layer, the system operationalizes the co-generation paradigm through two interconnected cycles known as the \textbf{Task Loop} and the \textbf{Interface Loop}. These loops are not independent processes but are rather sequential phases of a unified reaction cycle triggered by user behavior. The \texttt{ContextManager} acts as the central orchestrator that continuously listens to the input stream. It employs a trigger-based logic to determine when to invoke the agent's reasoning capabilities versus when to simply update the view. This aims that every generative action is grounded in the evolving intent of the user while maintaining a seamless and transparent collaboration. To elucidate this mechanism, we consider the scenario illustrated in Figure~\ref{fig:teaser} where a user interacts with a travel planning interface.

The cycle initiates when the system captures a user interaction. For instance, if the user selects the ``Accommodation'' module and applies a filter for ``Bed \& Breakfast'' (B\&B), this discrete manipulation is immediately logged into the \texttt{Bidirectional Action History}. This event acts as a trigger for the \textbf{Task Loop}. During this phase, the \texttt{TaskAgent} analyzes the newly updated history alongside the current \texttt{TaskDecomposition}. By observing that the user is editing the accommodation subtask with a specific preference for B\&Bs, the agent infers an implicit intent to refine the data scope. Consequently, the \texttt{TaskAgent} updates the subtask description and invokes the \texttt{ServiceAgent} to fetch relevant B\&B listings to populate the data layer. This inference process relies heavily on the temporal context provided by the history log, which allows the agent to distinguish between a casual exploration and a deliberate constraint change.

Once the \textbf{Task Loop} commits these semantic and data-level updates to the Context Layer, the \textbf{Interface Loop} is automatically triggered to synchronize the visual presentation. The \texttt{InterfaceAgent} retrieves the modified \texttt{TaskDecomposition} and synthesizes it with \texttt{Bidirectional Action History} to determine how these abstract changes should manifest visually. In our example, the agent recognizes that the underlying data now consists of B\&B listings rather than generic hotels. It generates an updated \texttt{InterfaceDescription} that might prioritize image-heavy cards suitable for vacation rentals over standard list views. Finally, the \texttt{RenderingAgent} translates this description into the live web interface. This completes the cycle as the user is presented with a new state ready for subsequent interaction. Through the continuous loop of these phases, DuetUI aims to make its interface more than just a static display; it seeks to let the interface dynamically reflect the shared understanding between humans and the agent.

\section{Technical Evaluation}

\label{sec:tech_evl}
We evaluate the system across diverse scenarios to ensure its robustness for the subsequent user study, while simultaneously examining how the co-generation mechanism specifically influences system performance.
\subsection{Experimental Setup}
We curated a dataset comprising 10 distinct tasks (see Appendix~\ref{section: task List}, which were adapted from prior research~\cite{yinOperationCognitionAutomatic2025, caoGenerativeMalleableUser2025, kimPlanTogetherFacilitatingAI2025, yenMemoletReifyingReuse2024, chenGapSynergyEnhancing2023} and refined based on our formative study) designed to ensure coverage of broader UI patterns. To simulate realistic usage variability and avoid generic outputs, we, referring to relevant works~\cite{xuCanLargeLanguage2024,barricelliEnduserDevelopmentEnduser2019a,huOSAgentsSurvey}, utilized GPT-4 to generate specific user personas for each task, assigning distinct demographic profiles (e.g., age, gender) and specific task-execution preferences. Furthermore, the test framework simulates user behavior based on these profiles, fine-tuned using user prompts and behavioral findings collected from our Formative Study. To establish a reliable benchmark, we invited a senior UX designer (E1) to annotate the Ground Truth for each scenario, providing both the ideal task description and a detailed interface specification.

To better understand the functional contribution of each architectural module, we also designed an ablation-style comparative experiment involving three conditions:
\begin{itemize}
\item \textbf{Baseline (GPT-4):} A standard direct prompting approach without intermediate reasoning steps or architectural scaffolding.
\item \textbf{DuetUI (No Loop):} A linear generation pipeline that utilizes \textit{DuetUI}'s six-stage generation process (including all agents but excluding the context manager), operating as a single-pass mechanism without iterative feedback.
\item \textbf{DuetUI (Full Loop):} The complete system featuring the \textbf{Bidirectional Context Loop}. This condition enables iterative refinement, where the Agent's automated analysis of the task and user GUI operations are encoded as context to guide subsequent generation steps.
\end{itemize}

\subsection{Methodology and Metrics}
Based on the above experimental setup, we collected 30 final outputs produced by the three systems across the ten tasks. Each output includes the full system logs as well as the final generated task description and interface description, comprising a total of 197 logs, 30 tasks, 118 \textit{Subtasks}, 225 \textit{Data} items, 30 \textit{Navigation} items, 100 \textit{Pages}, and 227 \textit{Components}. Note that due to the inclusion of data from ablated system variants, the element counts across these hierarchical levels do not strictly correspond. We then performed the following analyses. Our evaluation employed a hybrid approach combining automated reference-based metrics and expert human assessment.

\subsubsection{Automated Evaluation}
Considering the inherent variability of generative design, where multiple valid solutions correspond to a single intent, and to mitigate the impact of Verbosity Bias \cite{saito2023verbositybiaspreferencelabeling}, we drew on relevant works \cite{yinOperationCognitionAutomatic2025, caoGenerativeMalleableUser2025} and adopted \textit{functional equivalence matching} instead of rigid string matching. A generated item was considered a match if it achieved either (1) \textit{Information Equivalence} (conveying the same core information) or (2) \textit{Operational Equivalence} (enabling the same user actions). Since this required scoring a large number of items, and following prior work~\cite{yinOperationCognitionAutomatic2025,gebreegziabherMetricMateInteractiveTool2025,luMistyUIPrototyping2025}, we used GPT-4 as an automated judge to score each item twice according to the above criteria, with the first and third authors independently reviewing the results. We assessed generated items against the GT across hierarchical levels: \textit{Task}, \textit{Subtask}, \textit{Data}, \textit{Navigation}, \textit{Page}, and \textit{Component}. We calculated Precision, Recall, F1 scores, and a Weighted F1 (W-F1) averaged across levels. Inter-Rater Reliability (IRR) between the two independent LLM evaluators was verified using Cohen's Kappa.

\subsubsection{Expert Evaluation}
We recruited two domain experts (E2, Senior UX Designer; E3, Frontend Engineer) with over five years of industry experience. E2 was distinct from the GT annotator to minimize bias. The two experts and the first author held a calibration meeting, where several representative examples from the Formative Study were jointly scored to unify evaluation standards. Experts then evaluated outputs on a 5-point Likert scale across five dimensions adapted from prior work~\cite{caoGenerativeMalleableUser2025,chenGenerativeInterfacesLanguage2025a}: \textit{Efficiency}, \textit{Layout}, \textit{Completeness}, \textit{Aesthetics}, and \textit{Matching} (alignment with user intent). The \textit{Overall} score is reported as the equal-weighted average of these dimensions. As with the automated scoring, we computed IRR based on the experts' scores to ensure reliability. After the scoring, the first author also conducted open-ended interviews with each expert to gather their qualitative impressions of the system.

\subsection{Results}
\subsubsection{Ground-Truth-Based Analysis}
\label{subsec:ground}
Table~\ref{tab:technical_comprehensive} presents the evaluation results relative to the ground truth. \textit{DuetUI (Full)} achieved the highest aggregate performance with a W-F1 of \textbf{0.508}, outperforming both the Baseline (0.277) and \textit{DuetUI (No Loop)} (0.459). The high IRR scores for automated judging ($\kappa > 0.95$, details in Table~\ref{tab:technical_comprehensive}) confirm reliability. A detailed inspection of the Precision–Recall trade-off reveals distinct behaviors. While \textit{DuetUI (Full)} demonstrates superior Recall (e.g., 0.898 at the Subtask level), its Precision is lower than the \textit{No Loop} condition at granular levels (Data and Component). For instance, at the Data level, \textit{DuetUI (Full)} records a Precision of 0.318 versus 0.435 for \textit{No Loop}, despite significantly higher Recall (0.466 vs. 0.312). This pattern—higher Recall but lower Precision—suggests that, when the loop is enabled, the system may generate additional items not present in the original ground truth but still aligned with user needs. This tendency is more pronounced at finer-grained levels. We further examine this phenomenon in the Discussion (Section~\ref{sec:loop}), along with reflections on LLM-based evaluation methods (Section~\ref{sec:mock}).


\begin{table*}[t]
\centering
\setlength{\tabcolsep}{4pt}
\renewcommand{\arraystretch}{1.35}
\Description{The table reports detail extraction performance across three conditions—Baseline, DuetUI (No Loop), and DuetUI (Full). The first column lists the condition, the second shows the aggregated Weighted F1, and the remaining groups of columns list precision, recall, and F1 at four hierarchical levels: Subtask, Data, Page, and Component. DuetUI (Full) achieves the highest Weighted F1 of 0.508, improving over DuetUI (No Loop) and the Baseline. It also yields the best F1 at all four levels. The final row shows very high inter-rater reliability, with Cohen’s kappa ranging from 0.954 to 0.998 across levels.}
\caption{Details Precision (P), Recall (R), and F1 across four hierarchical levels.
\textbf{W-F1} denotes the aggregated weighted F1 score. Bold indicates the best performance.}
\label{tab:technical_comprehensive}

\begin{tabular*}{\textwidth}{@{\extracolsep{\fill}} l c
ccc ccc ccc ccc}
\hline
\multirow{2}{*}{\textbf{Condition}}
& \multirow{2}{*}{\textbf{W-F1}}
& \multicolumn{3}{c}{\textbf{Subtask Level}}
& \multicolumn{3}{c}{\textbf{Data Level}}
& \multicolumn{3}{c}{\textbf{Page Level}}
& \multicolumn{3}{c}{\textbf{Component Level}} \\
\cline{3-14}

& 
& \textbf{P} & \textbf{R} & \textbf{F1}
& \textbf{P} & \textbf{R} & \textbf{F1}
& \textbf{P} & \textbf{R} & \textbf{F1}
& \textbf{P} & \textbf{R} & \textbf{F1} \\
\hline

Baseline
& 0.277
& 0.568 & 0.485 & 0.518
& 0.189 & 0.151 & 0.161
& 0.450 & 0.285 & 0.315
& 0.139 & 0.106 & 0.116 \\

DuetUI (No Loop)
& 0.459
& \textbf{0.759} & 0.858 & 0.766
& \textbf{0.435} & 0.312 & 0.344
& \textbf{0.680} & 0.422 & 0.448
& \textbf{0.339} & 0.214 & 0.248 \\

\textbf{DuetUI (Full)}
& \textbf{0.508}
& 0.743 & \textbf{0.898} & \textbf{0.798}
& 0.318 & \textbf{0.466} & \textbf{0.367}
& 0.596 & \textbf{0.575} & \textbf{0.541}
& 0.285 & \textbf{0.436} & \textbf{0.325} \\

\hline
\textit{IRR ($\kappa$)}
& \textit{--}
& \multicolumn{3}{c}{\textit{0.997}}
& \multicolumn{3}{c}{\textit{0.958}}
& \multicolumn{3}{c}{\textit{0.998}}
& \multicolumn{3}{c}{\textit{0.954}} \\
\hline
\end{tabular*}
\end{table*}

\subsubsection{Expert Quality Assessment}
\label{subsec:expert}
As shown in Table~\ref{tab:likert_distribution}, \textit{DuetUI (Full)} achieved the highest ratings with an \textbf{Overall} score of \textbf{4.11 ($SD=0.36$)}. Statistical analysis indicates significant improvements for \textit{DuetUI (Full)} over the Baseline, particularly in \textit{Completeness} (Mann-Whitney U=10.5, $p=0.002$) and \textit{Overall} quality (t-test $t=-2.96$, $p=0.008$). Although the difference between \textit{No Loop} and \textit{Full Loop} was not statistically significant, the score distribution reveals a distinct positive shift, with \textit{DuetUI (Full)} receiving predominantly ``4'' and ``5'' ratings. Importantly, the \textit{Matching} score for \textit{DuetUI (Full)} was the highest (4.40), contradicting the drop in Precision. This suggests that the generated elements that did not match the GT were nonetheless perceived by experts as aligned with user needs. The experts' IRR score was 0.74, which we consider reliable relative to comparable HCI studies~\cite{caoGenerativeMalleableUser2025,mcdonaldReliabilityInterraterReliability2019}. Feedback from expert E2 during interviews further highlighted this advantage. In a travel planning scenario, while the Baseline provided only simple text fields for destination entry, \textit{DuetUI (Full)} interpreted abstract constraints (e.g., ``avoiding crowds,'' ``safety concerns'') into concrete functional widgets such as dynamic filters, charts, and richer data fields (e.g., \textit{safety rating}). E3 added that some of these newly generated fields also included more complete logic, such as conditional visibility.

\definecolor{heat}{HTML}{4b90e2} 

\newcommand{\heatcell}[1]{%
  \ifnum#1=0  \cellcolor{heat!0}#1\else
  \ifnum#1<3  \cellcolor{heat!5}#1\else
  \ifnum#1<6  \cellcolor{heat!10}#1\else
  \ifnum#1<9  \cellcolor{heat!15}#1\else
  \ifnum#1<12 \cellcolor{heat!20}#1\else
  \ifnum#1<15 \cellcolor{heat!25}#1\else
  \ifnum#1<18 \cellcolor{heat!30}#1\else
               \cellcolor{heat!35}#1%
  \fi\fi\fi\fi\fi\fi\fi
}

\begin{table*}[t]
\centering
\setlength{\tabcolsep}{3pt}
\renewcommand{\arraystretch}{1.45}
\Description{The table shows the distribution of five-point Likert ratings for interface quality across three conditions: Baseline, DuetUI (No Loop), and DuetUI (Full). The rows list five dimensions—Efficiency, Layout, Completeness, Aesthetics, and Matching—followed by an Overall row. For each condition, the table reports the mean score with standard deviation and, in heatmap cells, the number of 20 participants choosing each score from 1 to 5. Ratings concentrate on 3–5, with almost no 1 or 2 selections. DuetUI (Full) achieves the highest mean on every dimension, with Overall means of 3.64 for Baseline, 3.80 for DuetUI (No Loop), and 4.11 for DuetUI (Full).}
\caption{Distribution of Likert scores with mean and standard deviation across three conditions.
Each cell reports the number of a given score.
Background encodes a heatmap, where darker cells indicate higher participant concentration.}
\label{tab:likert_distribution}
\begin{tabular*}{\textwidth}{%
l
c p{1.4em} p{1.4em} p{1.4em} p{1.4em} p{1.4em}
c p{1.4em} p{1.4em} p{1.4em} p{1.4em} p{1.4em}
c p{1.4em} p{1.4em} p{1.4em} p{1.4em} p{1.4em}
}
\hline
\multirow{2}{*}{\textbf{Dimension}} 
& \multicolumn{6}{c}{\textbf{Baseline}}
& \multicolumn{6}{c}{\textbf{DuetUI (No Loop)}}
& \multicolumn{6}{c}{\textbf{DuetUI (Full)}} \\
\cline{2-19}

& Mean & 1 & 2 & 3 & 4 & 5
& Mean & 1 & 2 & 3 & 4 & 5
& Mean & 1 & 2 & 3 & 4 & 5 \\
\hline

Efficiency
& 3.30 ($\pm$0.79) & \heatcell{0} & \heatcell{4} & \heatcell{6} & \heatcell{10} & \heatcell{0}
& 3.65 ($\pm$0.53) & \heatcell{0} & \heatcell{0} & \heatcell{8} & \heatcell{11} & \heatcell{1}
& \textbf{3.80 ($\pm$0.35)} & \heatcell{0} & \heatcell{0} & \heatcell{4} & \heatcell{16} & \heatcell{0} \\

Layout
& 3.75 ($\pm$0.35) & \heatcell{0} & \heatcell{0} & \heatcell{5} & \heatcell{15} & \heatcell{0}
& 3.85 ($\pm$0.34) & \heatcell{0} & \heatcell{0} & \heatcell{3} & \heatcell{17} & \heatcell{0}
& \textbf{4.10 ($\pm$0.46)} & \heatcell{0} & \heatcell{0} & \heatcell{2} & \heatcell{14} & \heatcell{4} \\

Completeness
& 3.15 ($\pm$0.47) & \heatcell{0} & \heatcell{1} & \heatcell{15} & \heatcell{4} & \heatcell{0}
& 3.65 ($\pm$0.58) & \heatcell{0} & \heatcell{0} & \heatcell{9} & \heatcell{9} & \heatcell{2}
& \textbf{4.15 ($\pm$0.58)} & \heatcell{0} & \heatcell{0} & \heatcell{3} & \heatcell{11} & \heatcell{6} \\

Aesthetics
& 3.90 ($\pm$0.21) & \heatcell{0} & \heatcell{0} & \heatcell{2} & \heatcell{18} & \heatcell{0}
& 3.90 ($\pm$0.39) & \heatcell{0} & \heatcell{0} & \heatcell{3} & \heatcell{16} & \heatcell{1}
& \textbf{4.10 ($\pm$0.39)} & \heatcell{0} & \heatcell{0} & \heatcell{1} & \heatcell{16} & \heatcell{3} \\

Matching
& 4.10 ($\pm$0.61) & \heatcell{0} & \heatcell{0} & \heatcell{3} & \heatcell{12} & \heatcell{5}
& 3.95 ($\pm$0.50) & \heatcell{0} & \heatcell{0} & \heatcell{3} & \heatcell{15} & \heatcell{2}
& \textbf{4.40 ($\pm$0.52)} & \heatcell{0} & \heatcell{0} & \heatcell{1} & \heatcell{10} & \heatcell{9} \\

\hline
\textbf{Overall}
& \multicolumn{6}{c}{\textbf{3.64 ($\pm$0.35)}}
& \multicolumn{6}{c}{\textbf{3.80 ($\pm$0.26)}}
& \multicolumn{6}{c}{\textbf{4.11 ($\pm$0.36)}} \\
\hline
\end{tabular*}
\end{table*}

\section{User Study}
To collect authentic user feedback, we conducted user experiments based on the technical evaluation.
We recruited \textbf{24 participants} through campus social media platforms. Participants (N=24, 12 female, aged 18--31) reported diverse prior experiences with AI tools and UI generation (see Appendix \ref{user study participant}). All participants completed an online consent form before the study and received compensation of \textbf{\$14}. 

\subsection{Study Design}
We employed a within-subjects study design to evaluate our prototype, \textbf{DuetUI}, against a baseline system, \textbf{Stitch}\footnote{\url{https://stitch.withgoogle.com}}. Stitch was selected as a suitable baseline because it integrates both conversational interaction and generative interface capabilities, aligning with the core functionalities of our system. While comparable research systems~\cite{caoGenerativeMalleableUser2025,wuAIChainsTransparent2022,yinOperationCognitionAutomatic2025} are unavailable for replication (some claim to be open-sourced but provide empty repositories~\cite{yinOperationCognitionAutomatic2025}, while others release only datasets~\cite{caoGenerativeMalleableUser2025}), existing commercial tools also present notable barriers for end-users. Code-centric LLMs (e.g., Gemini, v0) remain highly technical—requiring debugging skills—and slow, often taking minutes to generate results. Image-based generators (e.g., Midjourney) are largely uncontrollable, frequently producing outputs detached from functional UI requirements. In contrast, Stitch offers a more accessible alternative that reliably handles vague inputs and instantly generates interfaces without prompt engineering. However, it operates strictly through linear dialogue: it lacks any loop mechanism, cannot incorporate users' on-screen interactions as contextual signals, and does not surface the agent's internal reasoning or execution details as part of the context visible to the user.
The primary independent variable was the system used, and the order of exposure to the two systems was counterbalanced across all participants to mitigate learning effects.
The study materials included the same set of tasks in technical evaluation.
These tasks were designed to reflect complex, everyday scenarios that our formative study participants had identified as particularly time-consuming or difficult. 
To balance participant engagement with experimental control, we implemented a structured task selection process. Upon starting the study, each participant was asked to select one task of their own interest from the pool of ten. Following their selection, two additional tasks were randomly assigned from the remaining nine. Each participant performed their three selected tasks on both systems, resulting in a total of six task sessions per participant (3 tasks $\times$ 2 systems). This approach ensured that while participants were motivated by a task relevant to them, the overall task distribution remained controlled and comparable across the study. This experimental design is improved from related work, and we will discuss it further in Section~\ref{sec:evaluation}.

\subsection{Procedure}
The study was conducted online via video conferencing. Participants' screens and audio were recorded throughout the session. The procedure consisted of three stages:
\begin{itemize}
\item \textbf{Introduction (10 minutes):} Participants were briefed on the study goals and received a short tutorial.
\item \textbf{Task phase (45 minutes):} Participants completed six tasks, each followed by an evaluation questionnaire. We measured usability through the \textbf{System Usability Scale (SUS)}~\cite{brooke1996sus} and workload through the \textbf{NASA Task Load Index (NASA-TLX)}~\cite{hart1988development}. Inspired by prior work~\cite{caoGenerativeMalleableUser2025,yeInteractionIntelligenceDeep2025,geGenComUIExploringGenerative2025,minMalleableOverviewDetailInterfaces2025}, we also had participants rate their Task Satisfaction, Interface Satisfaction, and AI Satisfaction on 5-point Likert scales to measure their subjective experience (detailed in Appendix~\ref{section: questionnaire}).
\item \textbf{Interview (15 minutes):} A semi-structured interview was conducted to gather qualitative feedback.
\end{itemize}

\subsection{Data Collection and Analysis}
Twenty-four participants completed six tasks, yielding 144 valid questionnaire sets and 25.86 hours of screen and audio recordings. The third author subsequently organized and aligned the user behavior captured in the videos with the corresponding system logs. Building on this integrated dataset, we employed a mixed-methods approach to compare DuetUI with the baseline.
All questionnaire scales demonstrated high internal consistency (Cronbach's Alpha). For quantitative analysis, data normality was assessed using the Shapiro–Wilk test. Subsequently, we applied paired-samples t-tests for normally distributed data ($Q_{question}$: $M_{DuetUI}$ vs. $M_{Stitch}$; $p$=$p$-value, $d$=Cohen's d) and Wilcoxon signed-rank tests for non-normal or ordinal data ($Q_{question}$: $M_{DuetUI}$ vs. $M_{Stitch}$; $p$=$p$-value, $r$=Rank-biserial correlation).
For qualitative analysis, the first author transcribed all recorded materials, including system logs, screen-capture videos, and audio-recorded interview transcripts, then conducted open coding and thematic analysis. Codes and themes were iteratively refined with co-authors to ensure analytical rigor.

\section{Findings}
\subsection{Participants' Behavior Analysis}
\subsubsection{Quantitative Interaction Analysis}
To understand user behavior across the two conditions, we analyzed system logs, screen-capture videos, and the aligned user–system interaction records. Across all tasks, there was no significant difference in overall task duration between the baseline and DuetUI conditions (Time Duration: $M_D$=188.38 vs. $M_S$=195.38; $p$=.5545, $r$=.157). However, users in DuetUI spent less time on prompting actions, including typing or selecting dialog options (Prompt Duration: $M_D$=20.12s vs. $M_S$=49.37s; $p$=.0003, $r$=.962), and their first prompt was also shorter in duration (First Prompt Duration: $M_D$=25.08s vs. $M_S$=57.28s; $p$=.0140, $r$=.667). The average length of text prompts was lower as well (Prompt Length: $M_D$=24.92 vs. $M_S$=56.31; $p$=.0174, $d$=-1.171; Chinese character). In addition to text input, DuetUI users interacted more frequently with the tangible interface widgets(as mentioned in Section~\ref{feat:tangible_agency}), averaging 6.6 interactions per task compared to 1.2 interactions via the traditional dialog box. We further examined how participants interacted with the system across rounds of user input and agent response. DuetUI users completed more interaction rounds (Rounds: $M_D$=6.91 vs. $M_S$=2.36; $p$=.0001, $d$=3.157) and generated more discrete interaction inputs (Inputs/Clicks: $M_D$=11.40 vs. $M_S$=2.60; $p$=.0002, $r$=-1.002). These additional interactions occurred throughout the task process rather than clustering in a single stage.

Finally, we analyzed the distribution of time spent in the Task Loop versus the Agent Loop across the six stages of the framework (as mentioned in Section~\ref{feat:staged_generation}). Each stage is advanced through explicit user actions; within a stage, participants may engage in multiple interaction rounds or skip the stage. Stages that were not entered were excluded from the statistics. Accordingly, the stage-level analysis reflects the average time distribution over stages that were actually engaged.
Users tended to spend more time in the Agent Loop during earlier stages, such as \textit{Define} and \textit{Empathize}, and more time in the Task Loop in later stages such as \textit{Explore} and \textit{Refine}. For example, participants often allowed the Agent to generate initial structures before making direct adjustments themselves. In the final \textit{Duet} stage, time spent in both loops was more evenly distributed. These observed patterns reflect how participants allocated initiative across different phases of their tasks, independent of the number of interaction rounds or the overall task duration. Further discussion of these behaviors is provided in Section~\ref{sec:loop}.

\begin{figure}[h]
    \centering
    \includegraphics[width=0.95\linewidth]{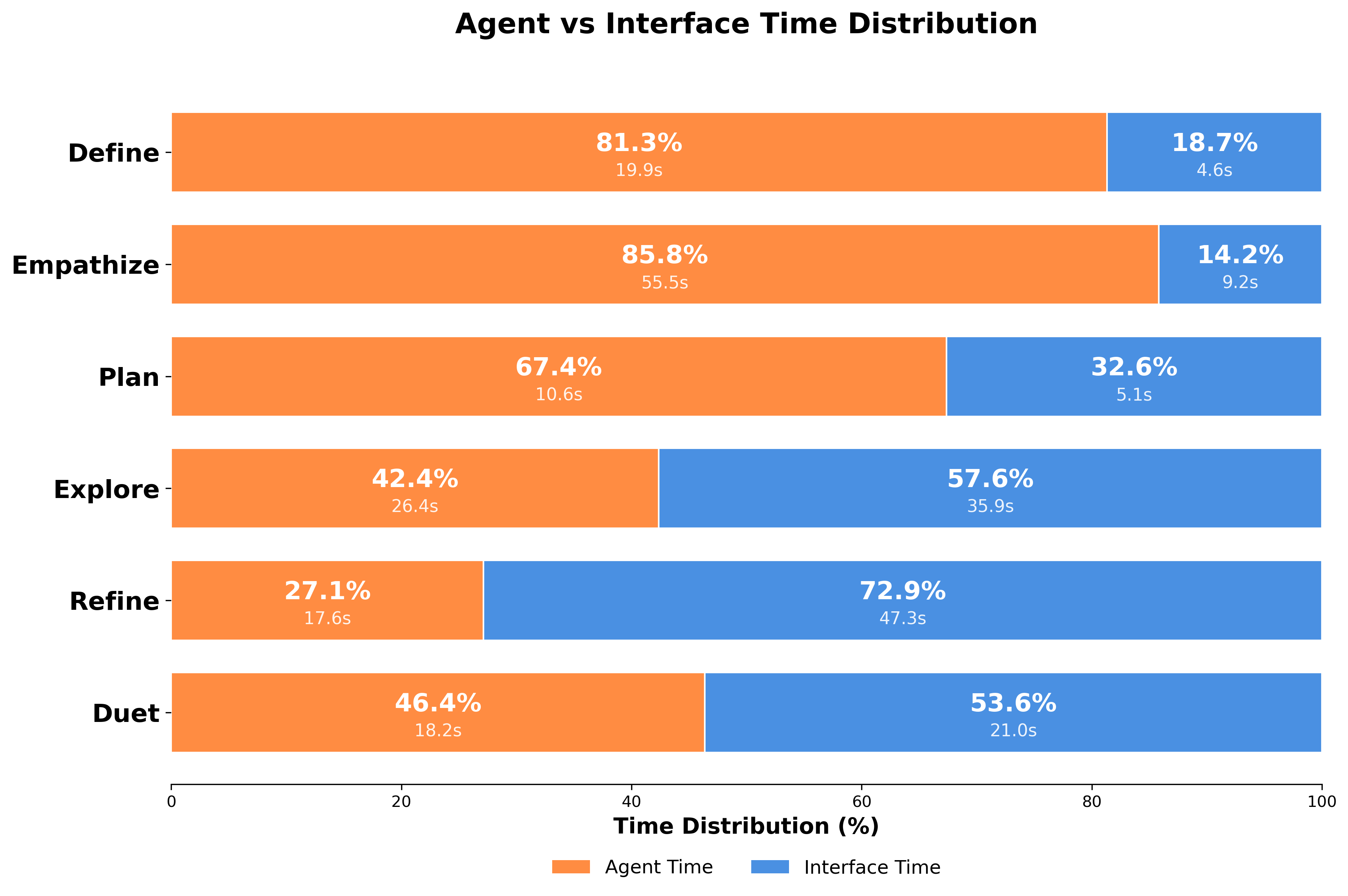}
    \caption{Visualization of user engagement in DuetUI (under a stage-level aggregation). Users shift from agent-led scaffolding to user-led refinement.}

    \Description{The Figure, a stacked horizontal bar chart titled "Agent vs Interface Time Distribution," which visualizes the time allocation (as percentages and raw seconds) between "Agent Time" (orange segments) and "Interface Time" (blue segments) across six sequential stages of the DuetUI co-generation process. For each stage (listed vertically on the left):Define: 81.3\% of time is Agent Time, with 18.7\% as Interface Time. Empathize: 85.8\% is Agent Time, 14.2\%  is Interface Time. Plan: 67.4\%  is Agent Time, 32.6\% is Interface Time. Explore: 42.4\% is Agent Time, 57.6\%  is Interface Time. Refine: 27.1\% is Agent Time, 72.9\%is Interface Time. Duet: 46.4\%  is Agent Time, 53.6\%  is Interface Time. The chart’s caption notes this is a "stage-level aggregation" showing a clear shift: early stages (Define, Empathize) are dominated by agent-led activity, while later stages (Explore, Refine) shift to user-led interaction via the interface—aligning with the process of moving from agent scaffolding to user refinement.}
    \label{fig:round_analysis}
\end{figure}

\subsubsection{Failure Analysis}
\label{subsec:failure}
Our failure analysis revealed several breakdown types. Five cases involved generation failure, where the system could not produce valid output even after three retries due to issues such as compilation errors, malformed JSON, or network instability. Other failures arose from unsupported user commands—for instance, P23's request for a fully customized UI style exceeded the capabilities of our renderer, which is tightly coupled with the Element Plus component library (Section~\ref{subsubsection: Agent Layer}), and P15 attempted to generate code dependent on external data sources the system cannot access. We also observed uses of the ``Regenerate Interface” button as a recovery mechanism: P16 changed their adoption target from dog to cat, and P6 switched travel mode from flying to train, both choosing explicit regeneration over conversational correction (as seen with P3–5, P12, P18, and P24). As P6 noted, they perceived the system as having ``failed” at that moment, suggesting that regeneration serves as a fallback when the conversational loop appears unstable. Participants additionally reported subtler loop-level issues—such as long latency caused by issuing multiple operations simultaneously (P7) or the agent missing nuanced intent (e.g., a preference for anaerobic exercise, P3)—which were often only partially articulated and therefore more difficult to detect.

\subsection{Self-Reported User Experiences}
\begin{figure}[ht!]
    \centering
    \includegraphics[width=0.95\linewidth]{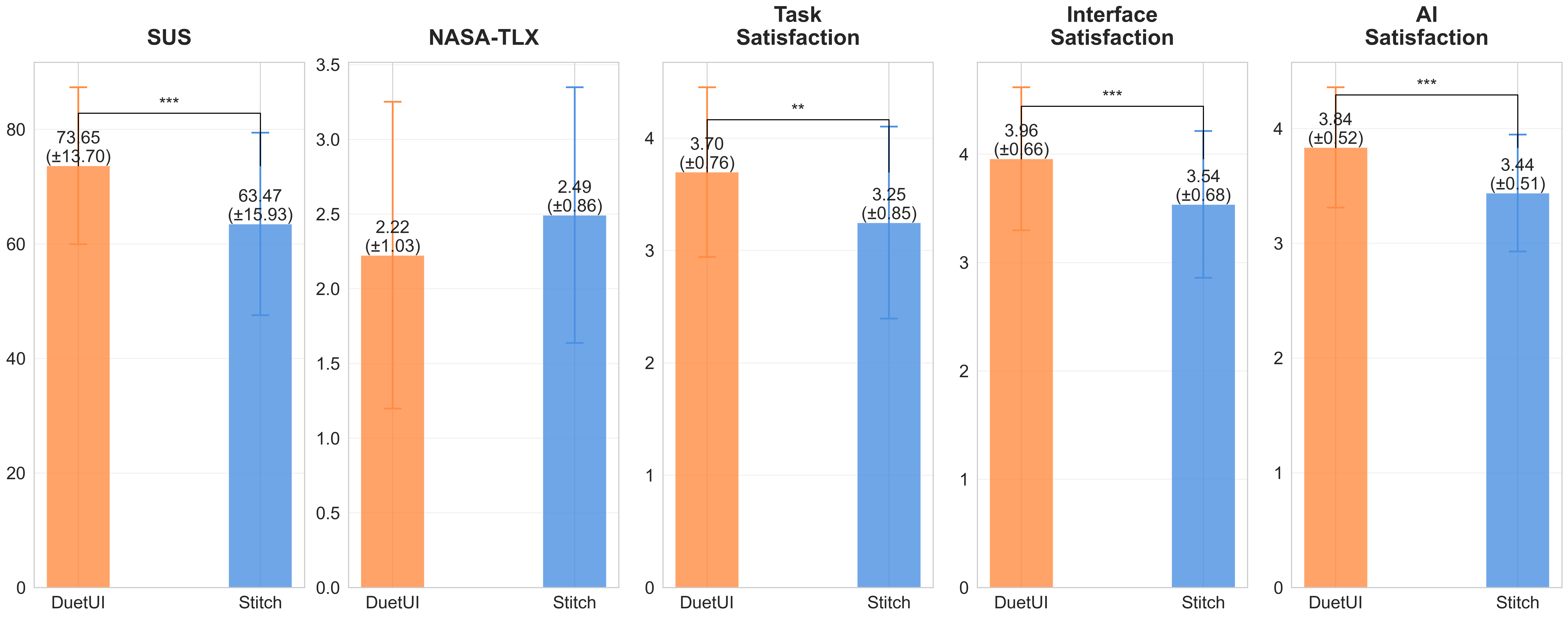}
    \caption{Subjective evaluation results comparing DuetUI and Stitch on usability (SUS), workload (NASA-TLX), Task Satisfaction, Interface Satisfaction, and AI Satisfaction. Bars show mean values and error bars represent standard deviation. (*** p < .001, ** p < .01)}
    \Description{Figure compares DuetUI and Stitch across five user-study metrics using side-by-side bar charts with error bars. From left to right, the panels show SUS, NASA-TLX, Task Satisfaction, Interface Satisfaction, and AI Satisfaction. In each panel, the orange bar represents DuetUI and the blue bar represents Stitch, annotated with the mean and standard deviation. DuetUI attains higher SUS scores (73.65 vs. 63.47), higher task, interface, and AI satisfaction (3.70, 3.96, and 3.84 vs. 3.25, 3.54, and 3.44), and slightly lower NASA-TLX workload (2.22 vs. 2.49). Asterisks between bars indicate statistically significant advantages for DuetUI on SUS, task satisfaction, interface satisfaction, and AI satisfaction, while NASA-TLX shows no significant difference.}
    \label{fig:all_metrics}
\end{figure}

Overall, our analysis revealed that, compared with the baseline system, DuetUI achieved higher usability, task satisfaction, interface satisfaction, and AI satisfaction, while maintaining a comparable level of cognitive workload, as shown in Figure \ref{fig:all_metrics}.

\begin{figure*}[p] 
    \centering  
    
    \begin{subfigure}{\linewidth}  
        \centering
        \includegraphics[width=0.88\linewidth]{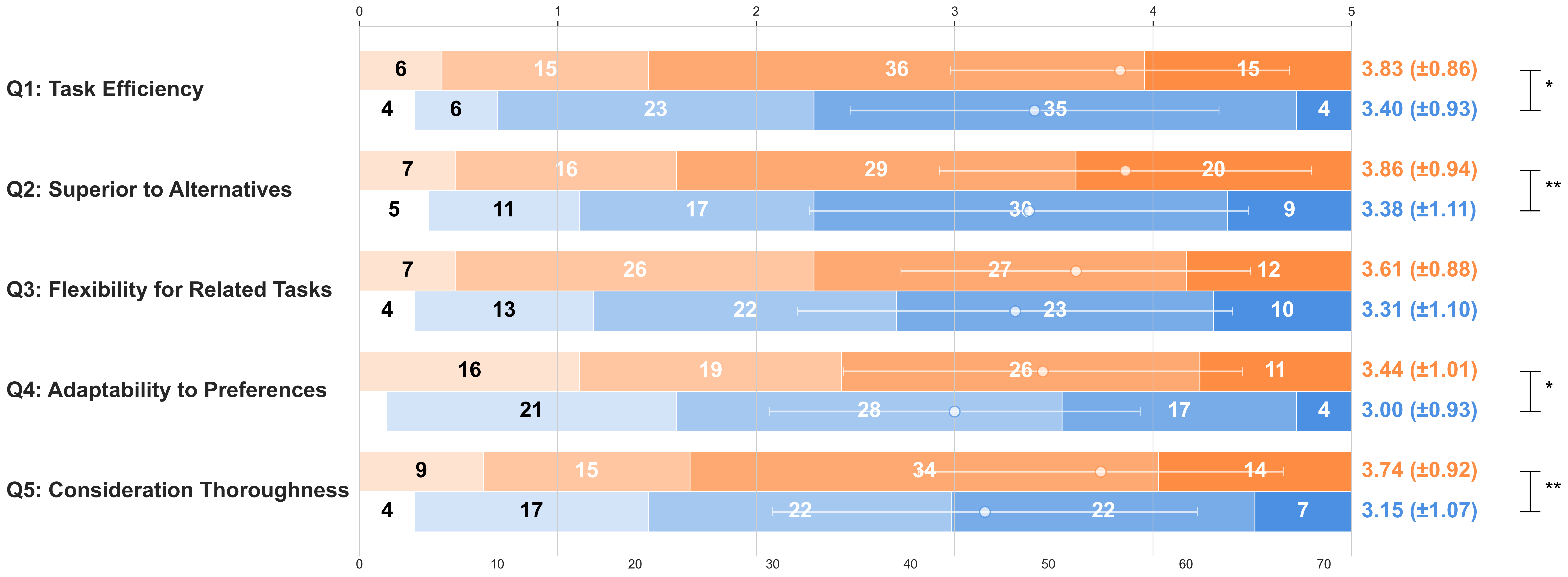} 
        \subcaption{Task Satisfaction} 
        \Description{Stacked bar chart comparing DuetUI and Stitch across five task satisfaction questions.}
        \label{subfig:task_satisfaction} 
    \end{subfigure}
    
    \begin{subfigure}{\linewidth}
        \centering
        \includegraphics[width=0.88\linewidth]{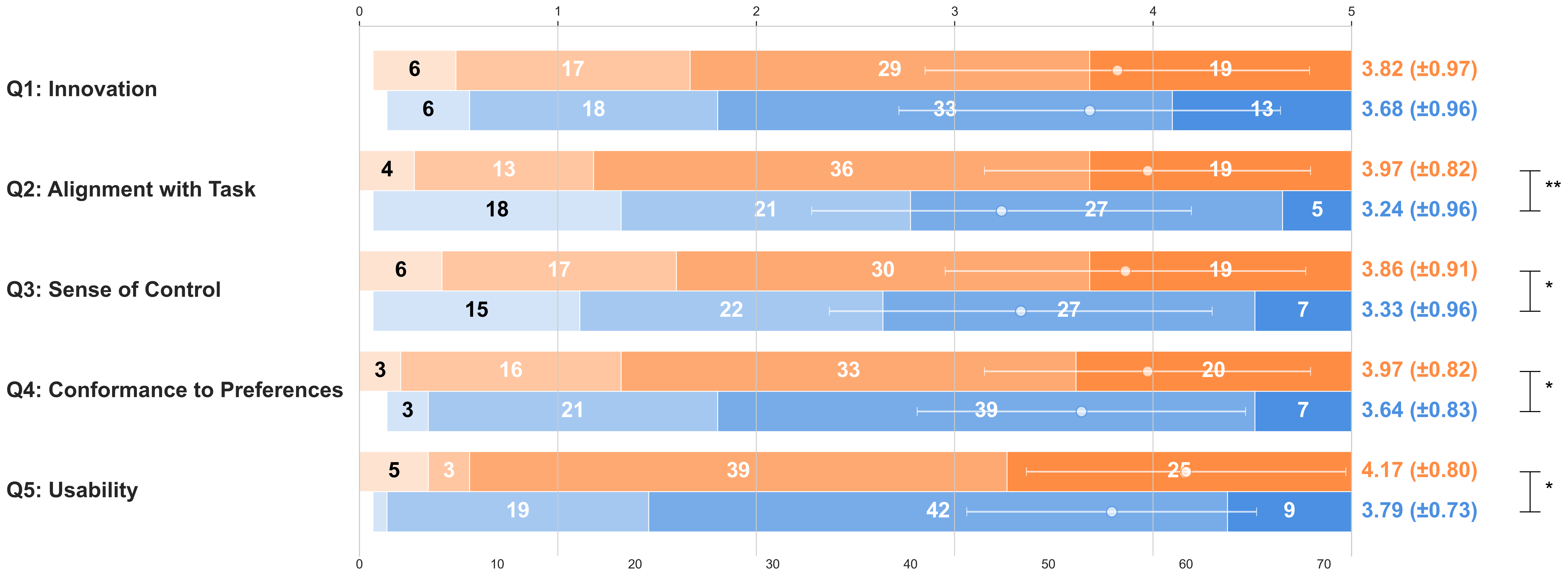}
        \subcaption{Interface Satisfaction} 
        \Description{Stacked bar chart comparing DuetUI and Stitch across five interface satisfaction questions.}
        \label{subfig:interface_satisfaction} 
    \end{subfigure}
    
    \begin{subfigure}{\linewidth}
        \centering
        \includegraphics[width=0.88\linewidth]{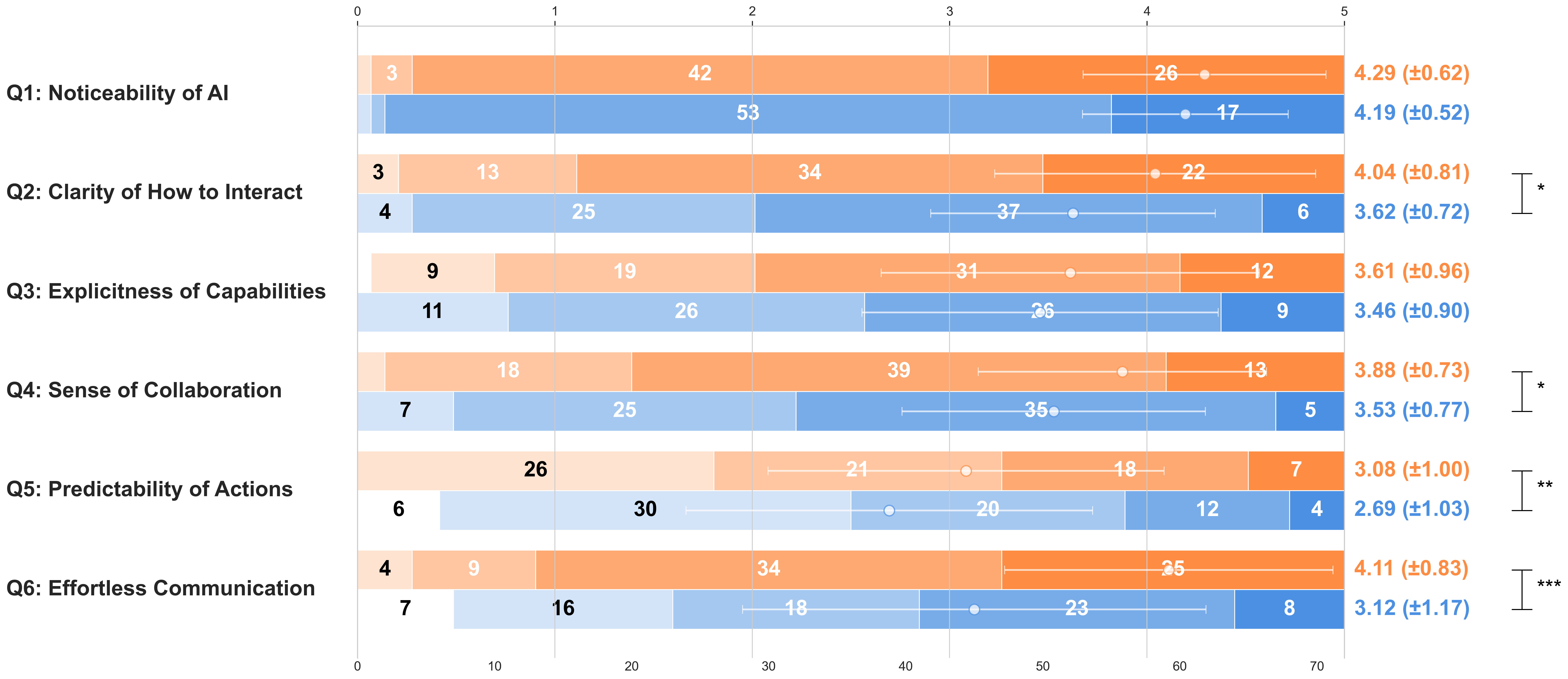}
        \subcaption{AI Satisfaction} 
        \Description{Stacked bar chart comparing DuetUI and Stitch across six AI satisfaction questions.}
        \label{subfig:ai_satisfaction} 
    \end{subfigure}
    \Description{Figure visualizes detailed Likert responses for DuetUI and Stitch across three groups of questions. Each panel uses horizontal stacked bar charts, with one bar per condition: an orange bar for DuetUI above a blue bar for Stitch. Within each bar, segments represent the number of participants selecting scores from 1 to 7, labeled along the top axis; segment lengths show the proportion of responses at each score. Panel (a), Task Satisfaction, includes five items: task efficiency, superiority to alternatives, flexibility for related tasks, adaptability to preferences, and consideration thoroughness. Panel (b), Interface Satisfaction, shows five items about innovation, alignment with task, sense of control, conformance to preferences, and usability. Panel (c), AI Satisfaction, lists six items: noticeability of AI, clarity of how to interact, explicitness of capabilities, sense of collaboration, predictability of actions, and effortless communication. To the right of each item, the mean rating and standard deviation for DuetUI and Stitch are displayed, with small brackets indicating statistically significant differences in favor of DuetUI on multiple questions. Overall, orange bars tend to shift toward higher scores compared with blue bars, illustrating consistently higher satisfaction ratings for DuetUI at both task and item levels.}
    \caption{User satisfaction ratings for DuetUI (orange) vs. Stitch (blue). Stacked bars show the score frequency distribution (bottom axis), while overlaid markers show the mean and standard deviation (top axis). Significance is marked as * p < .05, ** p < .01 and *** p < .001.} 
    \label{fig:satisfaction_distributions} 
\end{figure*}

\subsubsection{Higher Usability with a Lower Barrier to Entry}
DuetUI demonstrated significantly higher usability than the baseline (System Usability Scale: $M_{D}$=73.65 vs. $M_{S}$=63.5; $p$=.0019, $d$=.716), a sentiment shared by a majority of participants (87.5\%, 21/24). The main driver for this was its lower barrier to entry, which 20 users felt removed the need for specialized knowledge. Participants contrasted this with other tools that require ``\textit{prior AI design tool experience}'' (P14) or a technical understanding similar to ``\textit{taking a database class}'' (P6). Furthermore, eight participants (P1, P3, P6-7, P14-16, P18) praised the streamlined workflow from its ``\textit{all-in-one}'' nature, which saved them from having to ``\textit{repeatedly fill out forms or go to multiple apps}'' (P6) and ``\textit{do repetitive tasks like bookkeeping}''(P14). However, concerns about inclusivity remained: P4 cautioned that some specific user groups, such as older adults, might still face learning difficulties, remarking, \textit{``I'm worried my mom couldn't use this.''}

\subsubsection{Similar Workload but Improved Perceived Performance}
While the overall workload was statistically similar between the two systems (NASA-TLX: $M_D$=2.22 vs. $M_S$=2.49; $p$=.1061, $d$=-0.343), we observed a significant change in the composition of that workload. 
Participants reported that DuetUI required significantly less effort to achieve their goals ($Q_4$: $M_D$=2.32 vs. $M_S$=2.72; $p$=.0337, $d$=-.461). While other dimensions such as mental demand and frustration also showed lower mean scores for DuetUI, these differences did not reach statistical significance.
We speculate that DuetUI enabled users to shift their cognitive effort away from gathering information and managing AI communication, and toward the core task itself. This improved collaborative dynamic, which we believe explains the reduced performance demand, stems from the principles of a Clearer and More Controllable Agent Collaboration (Section \ref{sec:collaboration}), which we detail next.

\subsubsection{Greater Task Satisfaction and Efficiency}
DuetUI led to significantly higher overall task satisfaction (Task Satisfaction: $M_{D}$=3.7 vs. $M_{S}$=3.25; $p$=.0049, $d$=.635), detailed in Figure \ref{subfig:task_satisfaction}. 
Participants reported higher task satisfaction across several dimensions when using the proposed system. Specifically, they found it significantly more efficient ($Q_1$: $M_D$=3.83 vs. $M_S$=3.40; $p$=.0295, $d$=.474), and regarded it as a superior alternative to their usual phone usage patterns ($Q_2$: $M_D$=3.86 vs. $M_S$=3.38; $p$=.0067, $d$=.609). Furthermore, the system was perceived as more adaptable to their personal preferences ($Q_4$: $M_D$=3.44 vs. $M_S$=3.00; $p$=.0300, $d$=.472) and enabled a more thorough consideration of the tasks at hand ($Q_5$: $M_D$=3.74 vs. $M_S$=3.15; $p$=.0047, $d$=.639).
Supporting this, a vast majority of users (83.3\%, 20/24) participants would adopt DuetUI for daily use. Several (P1, P6, P12, P14, P17-18) noted it helped them be more comprehensive, with P1 adding, ``\textit{It helped me consider many task details I wasn't aware of... and some platforms I wouldn't normally think to use.}'' However, DuetUI was not universally suitable. Participants reported it could add complexity to simple tasks (P15), disrupt habits (P2, P7-8), and lack support for specialized professional tasks (P16). A key concern was the loss of community context by aggregating content. As P1 noted: ``\textit{The sense of community is quite important... If you mix them all up, I can't distinguish who posted it, or what kind of person they are.}''

\subsubsection{More Intelligent, Usable, and Aligned Interface}
Participants expressed significantly higher satisfaction with the interfaces generated by DuetUI
(Interface Satisfaction: $M_D$=3.96 vs. $M_S$=3.54; $p$=.0057, $d$=.622), detailed in Figure \ref{subfig:interface_satisfaction}. 
Participants found the interface to be significantly more aligned with their tasks ($Q_2$: $M_D$=3.97 vs. $M_S$=3.24; $p$=.0013, $d$=.749), with many describing its logic as clearer and more concise (P5-6, P10-12, P16, P22). The system also provided a stronger sense of control during interaction ($Q_3$: $M_D$=3.86 vs. $M_S$=3.33; $p$=.0149, $d$=.537). Furthermore, DuetUI's interfaces better conformed to user preferences ($Q_4$: $M_D$=3.97 vs. $M_S$=3.64; $p$=.0278, $d$=.480) and demonstrated significantly higher usability ($Q_5$: $M_D$=4.17 vs. $M_S$=3.79; $p$=.0100, $r$=.714).
Participants (P5, P6, P12) highlighted DuetUI's seamless link between interface generation and task completion. Unlike baseline systems that stop at static mockups before producing code, DuetUI leads directly to the final outcome, making it more effective. As P5 stated, DuetUI ``\textit{basically gives me the plan directly,}'' whereas the baseline ``\textit{just designs a program for you... [which] you [then use] to solve the problem.}'' However, this strong task-orientation was a drawback for more exploratory scenarios. Participants noted that phone use often lacks ``\textit{a very clear purpose}'' (P17), and felt DuetUI's ``\textit{forced sense of structure diminishes the joy of casually browsing}'' (P4).

\subsubsection{Clearer and More Controllable Agent Collaboration}
\label{sec:collaboration}
Participants reported significantly higher satisfaction with the AI agent in DuetUI (AI Satisfaction: $M_D$=3.84 vs. $M_S$=3.44; $p$=.0012, $d$=.752), detailed in Figure \ref{subfig:ai_satisfaction}.
Regarding AI satisfaction, participants found the interaction logic to be significantly clearer ($Q_2$: $M_D$=4.04 vs. $M_S$=3.62; $p$=.0128, $d$=.551) and experienced a stronger sense of collaboration with the system ($Q_4$: $M_D$=3.88 vs. $M_S$=3.53; $p$=.0162, $d$=.530). The AI's actions were perceived as more predictable ($Q_5$: $M_D$=3.08 vs. $M_S$=2.69; $p$=.0069, $d$=.606), and notably, DuetUI enabled significantly more effortless communication compared to the baseline ($Q_6$: $M_D$=4.11 vs. $M_S$=3.12; $p$=.0001, $d$=.946).
Qualitatively, 15 participants (P1, P3-6, P8-10, P12-18) stated that DuetUI effectively reduced their prompting burden. This was achieved by replacing single text prompts with clearer interaction affordances and by enabling real-time corrections. For example, P4 highlighted the value of structured inputs: ``\textit{For a non-developer like me... having a dropdown menu would be much better.}'' Participants also valued the ability to intervene when the AI went off-track, which was a common frustration with the baseline. P9 explained, ``\textit{I feel like I can correct the AI's mistakes in real-time, instead of waiting...}'' This reduced burden, as P4 added, ultimately ``\textit{increased their desire to interact with the AI.}''
Six participants (P2-4, P6, P10, P15) found DuetUI reduced their cognitive load by handling the burdensome process of ``\textit{gathering and integrating information}'' and providing ``\textit{suggestions},'' allowing them to ``\textit{refine the plan and focus on the task itself}'' (P3). P6 summarized this succinctly: ``\textit{This reduced my mental load significantly.}''

\subsection{User Perceptions of Human-Agent Co-Generation}
Our study provides qualitative insights into how end-users experience human-agent co-generation, which differs from professional tools for designers and developers, and identifies core design principles centered on collaborative process, the characteristics of the generated artifact, the dynamics of control, and the foundational element of trust.

\subsubsection{The Process: A Preference for Iterative Dialogue over One-Shot Generation}
A central finding is that participants overwhelmingly embraced the iterative nature of co-generation, valuing the journey of creation over a single, perfect output (P3, P5-7, P11-13, P17-18, P22, P24). This aligns with the core principle of our bidirectional context loop, which facilitates a continuous dialogue. The majority of these participants viewed the interaction as a collaborative dialogue, wanting to progressively refine the interface with the system. As P3 noted, the value lies in rapid iteration rather than initial perfection:
\begin{quote}
    \textit{``I don't expect it to get everything right in one go. I don't think that's realistic. The best thing is that it can iterate quickly.''} (P3)
\end{quote}
This sentiment was echoed by others who appreciated the feeling of being guided by a proactive partner rather than merely operating a passive tool, suggesting a desire for a turn-taking, collaborative model.
\begin{quote}
    \textit{``It's a step-by-step confirmation process, with guidance from the AI.''} (P4)
\end{quote}
However, this preference for guidance was not without its concerns. Some participants expressed unease about a perceived loss of freedom, feeling compelled to follow the AI's reasoning (P7, P15-16, P22). Others worried that the AI might misinterpret their intent and steer the design process in an undesirable direction (P7, P10).

\subsubsection{The Artifact: Valuing Predictability and Functional Coherence}
For the co-generated artifact to be effective, participants stressed two key qualities: predictability in its generation and logical coherence in its structure. Predictable AI behavior was seen as crucial for building a stable mental model of the system, making the partnership reliable. P14 lauded DuetUI for this stability in contrast to other generative tools:
\begin{quote}
    \textit{``The obvious difference is that DuetUI is very stable. No matter how it generates, it produces a generally consistent interface, whereas [another tool] is completely random.''} (P14)
\end{quote}
This consistency was framed as a necessary trade-off against excessive flexibility, which P1 found could be disorienting: \textit{``It would be quite unnerving if you saw a completely different interface every time you opened your phone.''} (P1)

Furthermore, when evaluating the final interfaces, participants consistently prioritized logical structure and usability over visual aesthetics. Seven participants specifically commended the interfaces from DuetUI for being logical and concise (P5-6, P10-12, P16, P22). As P12 articulated, functional clarity was the primary concern:
\begin{quote}
    \textit{``I didn't even notice the aesthetics... being simple and easy to use is the top priority.''} (P12)
\end{quote}
While utility was paramount, participants conceded that aesthetics could enhance long-term user retention, with P17 noting, \textit{``but being beautiful might make me use it a bit longer.''}

\subsubsection{The Control: A Need for Dynamic and Adaptive Autonomy}
~\label{sec:adaptive autonomy}
While DuetUI introduces a model for shared control, participant feedback highlighted that the ideal balance of control is not static but must be dynamic and adaptive. They indicated that the allocation of control should adjust to the task stage and, more importantly, to the user's evolving expertise. Users desired greater control over tasks they became more familiar with (P8, P14-15) and as their overall proficiency with the system increased (P9, P13). Participants also suggested that expert users, such as professional designers, should be granted more granular control from the outset, especially for specialized tasks (P6, P14, P23).

\subsubsection{The Foundation: Trust through Veracity and Transparency}
\label{sec:hallu}
Perhaps the most critical theme was the foundational role of trust in the human-agent partnership, a paramount concern for a majority of participants (P1, P3-4, P6-9, P11-18, P21). This apprehension was rooted in prior negative experiences with LLM ``\textit{hallucinations}'' (P4, P6-7, P13, P15-16, P18, P22) and in-study interactions where simulated data was perceived as factual errors. The fragility of this trust was powerfully articulated by P14:
\begin{quote}
    \textit{``The incorrect information directly led to the collapse of my trust in this system... and that trust is very hard to win back.''} (P14)
\end{quote}
Concerns also included the AI omitting critical information (P5-6), misinterpreting important details (P6), or using outdated data (P6, P8). In a positive finding, DuetUI's practice of citing data sources was noticed and appreciated (P1, P12, P14-15, P17), suggesting transparency is a direct pathway to building trust. To further bolster this foundation, participants proposed several mechanisms: providing direct links to original sources (P4, P6-7, P9, P11-12, P16-17), prioritizing official data sources (P14, P17), and implementing secondary verification agents (P13) or peer reviews (P8).

\section{Discussion and Future Work}
Our work on DuetUI serves as a proof-of-concept for a new paradigm of human–agent co-generation for interfaces. While our evaluation illustrates its immediate utility, it also reveals a broader design space and exposes several open challenges. In this section, we focus on the limitations of our current approach and outline key directions for future research, positioning DuetUI as an initial step rather than a complete solution toward more symbiotic human–agent interaction.

\subsection{Fostering Interface Alignment via Explicit and Implicit Contextual Loops}
\label{sec:loop}

In this work, we explore a new paradigm of human–AI co-generation for interfaces, where the system and user iteratively shape both the task and the interface itself. Iterative, progressive collaboration is well-studied~\cite{wuUICoderFinetuningLarge2024, zhang2024enhancinglanguagemodelrationality}, and approaches such as RLAIF~\cite{leeRLAIFVsRLHF2024} have demonstrated the benefits of agent-guided refinement. In the interface domain, similar ideas have been framed as \textit{malleable}~\cite{caoGenerativeMalleableUser2025, minMalleableOverviewDetailInterfaces2025} or \textit{evolutionary}~\cite{wangInteractionProcessInfrastructure2025} interfaces. Our work builds on these foundations by combining explicit user inputs with implicit signals extracted from user interactions, enabling collaboration across both the interface and task levels. Formative study findings (Section~\ref{sec:formative_study}) and objective user behavior data (Section 8.1) show that users engage in a dynamic, ascending loop, where early stages rely on the agent to clarify vague intentions and later stages shift control to the user for refinement. The context exchanged between user and agent is multi-layered: users provide explicit expressions of intent, while their interface interactions offer implicit behavioral cues. Our system translates these into an intermediate language between user and agent (Section~\ref{feat:tangible_agency}), and metrics indicate that users prefer this tangible approach over purely textual prompts. Implicit behavior inference reduces the need for explicit prompts: the system can deduce task goals from interface actions and highlight relevant UI components. Expert reviews (Section ~\ref{subsec:expert}) suggest that this design enriches the interaction experience and supports more detailed component-level design, helping bridge the Gulf of Execution~\cite{chenICardGenerativeAISupported2025, luMistyUIPrototyping2025}. 

We acknowledge that this approach may not be suitable for all types of tasks. Achieving effective human–AI symbiosis requires a delicate balance of initiative: overly helpful agents can lead to cognitive offloading and over-reliance~\cite{baumeisterEgoDepletionActive2018, vaccaroWhenCombinationsHumans2024}, whereas agents that are too proactive may feel paternalistic and stressful. Participants' behaviors reflected this tension; for example, P15 admitted, \textit{``I'll just pick one,''} when options were similar, while P12 appreciated a button that allowed them to \textit{``be lazy.''} P16 described overly proactive behaviors as an \textit{``impatient customer service agent''} pressuring them to make decisions. Our findings suggest that the ideal balance is dynamic, shifting with users' growing expertise (P9, P13) and task familiarity (P8, P14–15). This highlights the need for adaptive control, where agents can infer and adjust to users’ cognitive states and desired guidance in real-time.

Furthermore, a broader context may also introduces challenges. Our ablation experiments(Section ~\ref{sec:tech_evl}) showed a decrease in Precision, likely due to additional noise from more extensive context usage, a trend supported by several cases in our failure analysis(Section ~\ref{subsec:failure}). Integrating historical interactions and distilling a unified user model~\cite{xuCanLargeLanguage2024} may help mitigate this issue over long-term interaction. Overall, our system serves as a proof of concept, with substantial room for refinement before real-world deployment. We hope that these findings inform future research and industrial practice, and we outline several key open questions and limitations in the following sections.

\subsection{Evaluation Metrics and Methodologies Toward Real-World Generative Interfaces}
\label{sec:evaluation}
We assessed DuetUI using three complementary methods: ground-truth-based metrics, expert review, and user self-reports. This multi-dimensional approach revealed a critical tension: while \textit{DuetUI (Full)} improved automated Recall and Expert Matching (reported in Section~\ref{sec:tech_evl}), it exhibited a decrease in Precision compared to the \textit{No Loop} baseline. This reduction largely reflects the bidirectional loop generating functional widgets (e.g., dynamic filters) that were absent in the static Ground Truth. Traditional metrics penalize these valid enhancements as ``errors,'' indicating that proximity to a fixed Ground Truth is an insufficient proxy for user experience in generative contexts. Accordingly, we emphasize user-centered evaluation over purely automated frameworks, consistent with prior work on malleable interfaces~\cite{wuUICoderFinetuningLarge2024, chenGenerativeInterfacesLanguage2025a, caoGenerativeMalleableUser2025, minMalleableOverviewDetailInterfaces2025}. Our results underscore that the \textit{process} of generation can be as important as the final artifact, suggesting that future evaluation frameworks should integrate adaptive, context-aware metrics capable of capturing functional creativity beyond static references.

Regarding evaluation scope, we note a common limitation in generative UI research: many systems are implemented without preliminary user research~\cite{vaithilingamDynaVisDynamicallySynthesized2024, yeInteractionIntelligenceDeep2025, caoGenerativeMalleableUser2025, wuAIChainsTransparent2022}. Our formative study provided initial guidance, but it relied on only two task scenarios, which may limit generalizability. To broaden evaluation, we selected ten cross-application tasks combining existing datasets~\cite{yinOperationCognitionAutomatic2025, caoGenerativeMalleableUser2025, kimPlanTogetherFacilitatingAI2025, yenMemoletReifyingReuse2024, chenGapSynergyEnhancing2023} and our formative insights, aiming to cover tasks that single apps cannot fully address. Participants chose one task of personal interest alongside two random assignments, balancing motivation for realistic engagement with experimental comparability, following strategies reported in prior studies~\cite{caoGenerativeMalleableUser2025}. Other studies have adopted predefined task sets (e.g., ~\cite{yinOperationCognitionAutomatic2025} with ten tasks of varying difficulty and cognitive demands; ~\cite{chenGapSynergyEnhancing2023} with twelve campus scenarios), while some allowed free-form exploration~\cite{kimPlanTogetherFacilitatingAI2025}, which can reduce alignment between experimental conditions. Our approach seeks a middle ground that supports realistic user choices while maintaining baseline comparability. Future work could extend this by integrating larger, more diverse task sets, potentially across real-world applications, to better evaluate generalizability and robustness in authentic usage contexts.

Finally, we acknowledge the limitations of our sample size. Although our within-subjects cohort ($N=24$) exceeds typical generative interface studies (often 8–20 participants~\cite{yinOperationCognitionAutomatic2025, kimPlanTogetherFacilitatingAI2025, wuAIChainsTransparent2022, caoGenerativeMalleableUser2025}), it remains modest compared to large-scale quantitative surveys, which may provide broader demographic representativeness and additional power to explore individual differences. Nevertheless, this sample meets local HCI standards~\cite{caineLocalStandardsSample2016} and achieved robust post-hoc statistical power (SUS: $1-\beta = 0.919$; Satisfaction metrics: $>0.83$), supporting confidence in the observed effects as a foundation for future, larger-scale validations. Future studies could further increase participant diversity and sample size to capture a wider range of user behaviors, enabling more generalizable insights and informing scalable evaluation protocols for real-world generative interfaces.

\subsection{Navigating the Trade-offs Between Simulation and Reality in System Design}
\label{sec:mock}
To prioritize the evaluation of the interaction paradigm over backend instability, we adopted a simulation-based approach for external services~\cite{caoGenerativeMalleableUser2025}. We observe that current research often navigates data integration challenges by selecting tasks with minimal external dependencies~\cite{wuAIChainsTransparent2022,kimCellsGeneratorsLenses2023,yeInteractionIntelligenceDeep2025} or relying on manual user injection~\cite{kimPlanTogetherFacilitatingAI2025,vaithilingamDynaVisDynamicallySynthesized2024}. We infer that these methodological choices likely stem from systemic bottlenecks within the field. Despite the emergence of standards like MCP~\cite{hou2025modelcontextprotocolmcp}, our survey of API-based LLM tools suggests that state-of-the-art agents continue to face notable reliability issues in complex tool-use scenarios~\cite{zhangLoopToolClosingDataTraining2025,qin2023toolllm}. Even the strongest closed-source systems achieve less than 80\% accuracy in multi-turn interactions, and prior studies highlight persistent challenges in determining when to call an API, which API to select, and how to construct valid and effective API requests~\cite{zhangLoopToolClosingDataTraining2025}. By utilizing high-fidelity mock data~\cite{caoGenerativeMalleableUser2025}, we effectively reduced the friction caused by unstable real-world APIs to focus our analysis on the human-agent co-generation process. However, our study reveals that simulation alone does not resolve the ``trust gap'' as participants noted (detail in Section~\ref{sec:hallu}) that hallucinations persist regardless of the data source. Consequently, we suggest that future deployments prioritize the development of interactive mechanisms for real-time source verification and iterative error repair to establish long-term user trust.

Regarding technical evaluation, we leveraged synthetic user inputs and LLM-as-a-judge methodologies~\cite{gebreegziabherMetricMateInteractiveTool2025}, a choice that aligns with growing trends in recent HCI literature~\cite{chenDesignGuidelineRPA2025a,caoGenerativeMalleableUser2025,xuCanLargeLanguage2024,huOSAgentsSurvey}. While Morris characterizes synthetic data as ``crucial yet controversial''~\cite{morrisHCIAGI2025}, we argue that it offers essential scalability and safety for testing early-stage systems, particularly given the significant data disparity between academia and industry~\cite{barricelliEnduserDevelopmentEnduser2019a}. This contrast is evident when comparing our constraints to industry players who can leverage massive behavioral logs, such as the 200k sessions reported by Tomlinson et al.~\cite{tomlinsonWorkingAIMeasuring2025}. Academia currently lacks comparable large-scale datasets that specifically capture the collaboration between ordinary users and generative interfaces. Therefore, we identify the construction of such open, high-quality interaction datasets as a vital future direction to bridge the divide between simulation-based experiments and robust, real-world applications.

\subsection{Beyond the GUI: Richer Context and Multimodal Interaction}
A truly proactive partner requires a richer contextual understanding. Future iterations may incorporate real-time sensor data, such as GPS or physical activity, to infer the user's immediate situation~\cite{yangContextAgentContextAwareProactive2025, chenGapSynergyEnhancing2023}. Moreover, establishing a persistent memory pool to distill long-term user profiles from historical interactions would enable the agent to cold-start new tasks. This evolution allows the system to offer more holistic, personalized support tailored to the user's habits~\cite{zhangEmpoweringAgileBasedGenerative2024, xuCanLargeLanguage2024}. Furthermore, the principles of our co-generation paradigm are fundamentally modality-agnostic, extending far beyond the GUI. Grounded in Multiple Resource Theory~\cite{Wickens_2002}, the agent can act as an intelligent resource allocator that distributes information across cognitive channels to reduce workload. For instance, the system could dynamically generate voice interfaces upon detecting cognitive strain~\cite{adaptivevox, Khawaja_Chen_Marcus_2014}, or co-create layouts for emerging platforms like AR~\cite{niyazovUserDrivenConstraintsLayout2023} and human-robot interaction~\cite{geGenComUIExploringGenerative2025}. Ultimately, integrating Ability-Based Design principles~\cite{wobbrockAbilityBasedDesignConcept2011, wobbrockAbilityBasedDesign2018} points toward a future where interfaces are co-generated not just for the task, but for the unique abilities of each user.

\section{Conclusion}
This paper challenges the prevailing agent-centric model of task automation by proposing and validating a new paradigm: human-agent co-generation. Embodied in our system, DuetUI, this paradigm envisions the human-computer relationship from one of command-and-control to a symbiotic partnership. The engine of this partnership is the bidirectional context loop, a novel mechanism where the agent scaffolds the task and the user's direct manipulations implicitly steer the collaborative process. Our technical ablation study and user study demonstrated that this approach not only significantly improves task efficiency and interface usability but, more importantly, fosters a transparent and fluid collaboration that better aligns with evolving human intent. By treating the interface as a shared, malleable artifact co-created by both human and agent, our work offers a promising design pattern for the future of interactive AI, moving beyond mere automation toward to man-computer symbiosis.

\section{Disclosure about Use of LLM}
This work involved the use of several large language models, including GPT-4, GPT-4o, Groq LLaMA3-70B, Gemini-2.5-pro, and Claude-4-sonnet. Specifically, GPT-4, GPT-4o, and Groq LLaMA3-70B were emoloyed as system component generation, LLM-as-judge annotation, and dataset synthesis; Gemini-2.5-pro and dev0 were employed in probe systems; Claude-4-sonnet was primarily employed for debugging; and Gemini-2.5-pro was additionally used for language refinement. 



\begin{acks}
We acknowledge the support of the Guangdong Provincial Key Laboratory of Integrated Communication, Sensing and Computation for Ubiquitous Internet of Things (Grant No. 2023B1212010007) and the 111 Center (Grant No. D25008).
This work was completed during the internships of Shaowen Xiang and Ruoting Sun in the Human-Centered Intelligence+ Summer Research Program, and we sincerely thank the program and support team.
\end{acks}

\bibliographystyle{ACM-Reference-Format}
\bibliography{chi26-175}


\onecolumn
\appendix
\section{Formative Study}

\subsection{Probe}
The probe system was implemented as a web application using the React JavaScript library~\footnote{\url{https://react.dev}}. It was deployed on a server, allowing users to access it remotely via a web browser.

\label{section: probe}
\begin{figure}[H]
    \centering
    \includegraphics[width=0.85\linewidth]{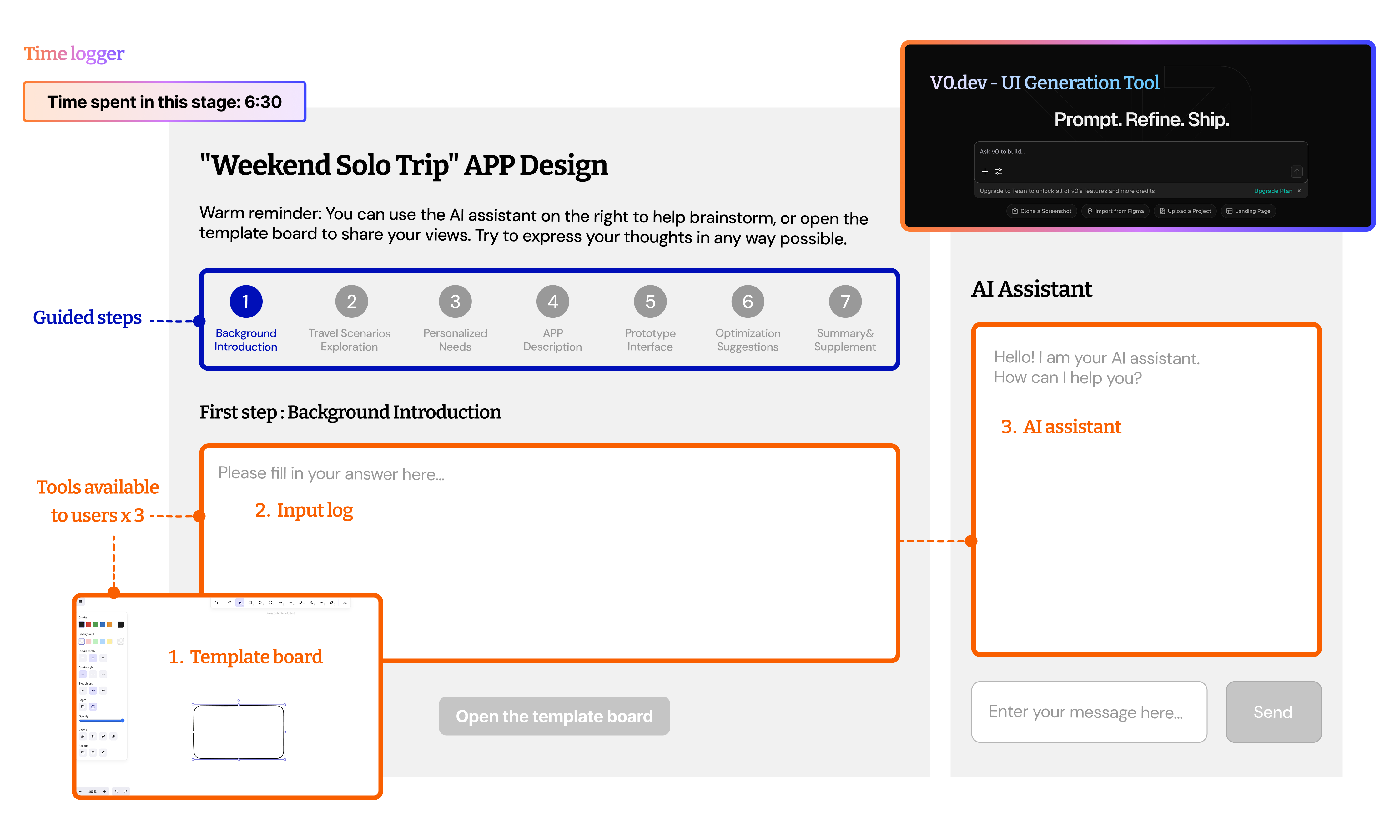}
    \caption{Screenshot of a probe system.}
    \Description{The figure shows the probe interface used in the formative study for the “Weekend Solo Trip” app design task. At the top left, a time logger displays “Time spent in this stage: 6:30,” indicating how long the participant has been in the current step. Below, the main workspace is split into two columns. The left column contains the guided workflow for seven steps, with step 1, Background Introduction, highlighted in a horizontal progress bar that also lists later stages such as Travel Scenarios Exploration, Personalized Needs, and APP Description. Under the heading “First step: Background Introduction,” three tools available to participants are outlined in orange: a template whiteboard at the bottom left for sketching ideas, a large input log text area in the center where participants can type free-form descriptions, and, on the right column, an AI Assistant chat panel that starts with “Hello! I am your AI assistant. How can I help you?”. At the top right, a screenshot of the V0.dev UI Generation Tool suggests an example external design tool participants may reference.}
    \label{fig:probe}
\end{figure}

\subsection{Procedure}
\begin{figure}[H]
    \centering
    \includegraphics[width=0.85\linewidth]{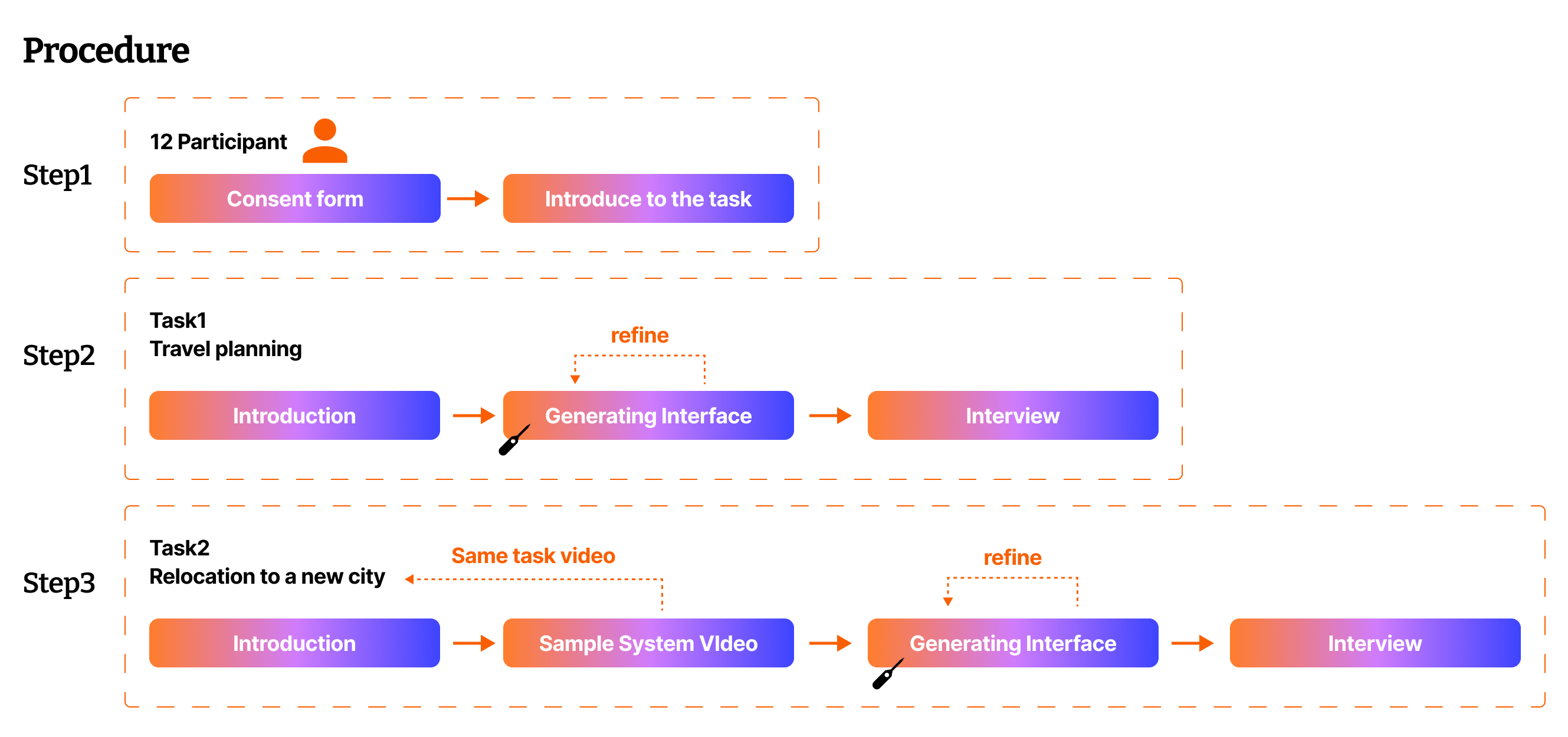}
    \caption{Diagram of the study procedure}
    \Description{The figure illustrates the procedure of the summative study as a three-step flow diagram. Step 1 shows 12 participants completing a consent form and then being introduced to the task. Step 2, labeled Task 1: Travel planning, contains a sequence of colored bars: Introduction, Generating Interface, and Interview. A dotted arrow labeled “refine” loops over the Generating Interface bar, indicating iterative design. Step 3, Task 2: Relocation to a new city repeats a similar sequence: Introduction, Sample System Video, Generating Interface, and Interview. A dotted label “Same task video” points to the Sample System Video, and another “refine” loop appears over the Generating Interface.}
    \label{fig:Procedure}
\end{figure}
\section{User Study}

\subsection{Participant Information}
\label{user study participant}
The following table presents the demographic data and prior AI tool usage for the 24 participants. Before the main study, they were asked to respond to the prompt, \textit{"Which of the following AI tools or features have you used before?"}, by selecting all applicable options.

The options provided in the survey were:
\begin{itemize}
    \item \textbf{Chatbots} (e.g., ChatGPT, Gemini, DeepSeek)
    \item \textbf{AI-assisted Code Generation} (e.g., GitHub Copilot, Cursor)
    \item \textbf{AI-assisted Design Tools} (e.g., Figma AI, Adobe Firefly)
    \item \textbf{Text-to-Image Generation} (e.g., Midjourney, Stable Diffusion)
\end{itemize}

Table \ref{tab:participant_info} summarizes their responses.

\begin{table*}[h]
  \centering
  \caption{Demographics and prior AI tool usage of participants. A checkmark (\checkmark) indicates the participant has used the corresponding tool.}
  \label{tab:participant_info}
    \begin{tabular}{l c c c c c c}
    \toprule
    \textbf{Participant} & \textbf{Age} & \textbf{Gender} & \textbf{Chatbot} & \textbf{Code Gen.} & \textbf{Design} & \textbf{Text-to-Image} \\
    \midrule
        P1  & 24 & Male   & \checkmark &            &            &                \\ 
        P2  & 22 & Female & \checkmark &            &            &                \\ 
        P3  & 22 & Female & \checkmark &            &            &                \\ 
        P4  & 22 & Male   & \checkmark & \checkmark &            & \checkmark     \\ 
        P5  & 18 & Female & \checkmark &            & \checkmark &                \\ 
        P6  & 22 & Female & \checkmark & \checkmark & \checkmark & \checkmark     \\ 
        P7  & 23 & Male   & \checkmark & \checkmark &            & \checkmark     \\ 
        P8  & 27 & Male   & \checkmark &            & \checkmark &                \\ 
        P9  & 27 & Female & \checkmark &            &            &                \\ 
        P10 & 19 & Male   & \checkmark & \checkmark &            & \checkmark     \\ 
        P11 & 19 & Male   & \checkmark & \checkmark &            & \checkmark     \\ 
        P12 & 26 & Female & \checkmark & \checkmark &            &                \\ 
        P13 & 25 & Male   & \checkmark & \checkmark &            & \checkmark     \\ 
        P14 & 23 & Female &            &            & \checkmark & \checkmark     \\ 
        P15 & 24 & Female & \checkmark & \checkmark & \checkmark & \checkmark     \\ 
        P16 & 26 & Female & \checkmark &            & \checkmark & \checkmark     \\ 
        P17 & 26 & Female & \checkmark & \checkmark & \checkmark &                \\ 
        P18 & 22 & Female &            &            &            &                \\ 
        P19 & 21 & Male   &            &            &            &                \\ 
        P20 & 19 & Male   & \checkmark & \checkmark &            &                \\ 
        P21 & 31 & Male   & \checkmark & \checkmark & \checkmark & \checkmark     \\ 
        P22 & 24 & Male   &            &            &            &                \\ 
        P23 & 20 & Female & \checkmark &            & \checkmark & \checkmark     \\ 
        P24 & 30 & Male   &            &            &            &                \\ 
        \bottomrule
    \end{tabular}
    \Description{The table summarizes demographics and prior AI tool usage for 24 participants. Each row lists a participant ID, age, gender, and checkmarks indicating whether they have used four categories of AI tools: chatbots, code generation tools, design tools, and text-to-image tools. Participants range in age from 18 to 31 and include both male and female users. Most participants have prior experience with chatbots, while fewer have used code generation, design, or text-to-image systems. Several participants report no prior AI tool usage at all, providing a mix of novice and experienced users for the study.}
\end{table*}

\subsection{Questionnaire}
\label{section: questionnaire}
This appendix presents the three questionnaires employed in our study to evaluate the user experience. All items were rated on a 5-point Likert scale, ranging from 1 (Strongly Disagree) to 5 (Strongly Agree). The questionnaires are designed to assess Task Satisfaction, Interface Satisfaction, and AI Satisfaction, respectively.

\subsubsection*{Task Satisfaction Questionnaire (TSQ)}
\begin{enumerate}
    \item I perceive the system as highly efficient for accomplishing my task.
    \item I consider this system a better alternative to my usual method for completing this task.
    \item I believe the system is flexible enough to be adapted for other related tasks.
    \item I feel the system adapted well to my personal preferences and habits during the task.
    \item The system provided sufficient and helpful information, considering many details on my behalf.
\end{enumerate}

\subsubsection*{Interface Satisfaction Questionnaire (ISQ)}
\begin{enumerate}
    \item I perceive the user interface as novel.
    \item The design of the interface is well-aligned with the task workflow.
    \item The interface provides a strong sense of control; I clearly understood the outcome of my interactions.
    \item I feel that the interface is tailored to my personal preferences.
    \item I believe the user interface is easy to use.
\end{enumerate}

\subsubsection*{AI Satisfaction Questionnaire (ASQ)}
\begin{enumerate}
    \item I was clearly aware of the AI's assistance while using the system.
    \item It was clear to me how to interact with the AI.
    \item I have a clear understanding of the AI's capabilities and limitations.
    \item I felt a sense of close collaboration with the AI while completing the task.
    \item I found the AI's behavior to be predictable; I was not concerned about it performing unexpected actions.
    \item I found communicating with the AI to be effortless.
\end{enumerate}

\subsection{Task List}
\label{section: task List}
\begin{table*}[h!]
\centering
\caption{Task descriptions.}
\label{tab:task_list_simple}
\begin{tabular}{p{0.20\linewidth} p{0.75\linewidth}}
    \toprule
    \textbf{Task Name} & \textbf{Description} \\
    \midrule
Solo Weekend Trip & Plan a complete one-day weekend trip including destination guide, transport, dining, and accommodation arrangements. \\
\addlinespace
Interview Preparation & Provide comprehensive interview preparation, covering company background, interview experience, and position requirements. \\
\addlinespace
Team Building Activity & Plan a corporate team-building event, including location selection, procurement of necessary materials, and overall activity design. \\
\addlinespace
Mobile Price Comparison & Perform multi-platform mobile phone price comparison, including specifications analysis and purchasing recommendation. \\
\addlinespace
Learning a Handicraft Skill & Design a full learning path, including tutorials, purchasing tools, and joining communities of interest. \\
\addlinespace
Monthly Budget Analysis & Analyze monthly expenditures from multiple platforms to provide financial recommendations and budgeting advice. \\
\addlinespace
Fitness Plan & Develop personalized fitness plans including gym/coach selection, diet plans, and purchase of exercise equipment. \\
\addlinespace
Home Renovation Plan & Plan home improvement projects, including design inspiration, product selection, and implementation strategies. \\
\addlinespace
Pet Adoption & Prepare for pet adoption and care, including finding adoption channels, preparing basic supplies, and learning care tips. \\
\addlinespace
Moving Plan & Plan the relocation process including housing search, moving services, and settling into a new environment. \\
\bottomrule
\end{tabular}
\Description{The table lists ten task scenarios used in the study and provides a short description for each. The first column gives the task name, and the second column explains what the user is asked to plan or analyze. Tasks span both personal and work contexts, including planning a solo weekend trip, preparing for a job interview, organizing a team-building activity, comparing mobile phone prices, learning a handicraft skill, conducting monthly budget analysis, creating a fitness plan, designing a home renovation plan, preparing for pet adoption, and planning a moving process.}
\end{table*}

\end{document}